\documentclass[
  a4paper, %
  10pt, %
  twoside, %
]{LTJournalArticle}

\usepackage[british]{babel}
\usepackage{subfig}
\runninghead{Patient-specific prediction of glioblastoma growth} %

\footertext{Submitted --- Under review} %

\title{Patient-specific prediction of glioblastoma growth via reduced order modeling and neural networks} %

\author{%
  D. Cerrone$^{1}$ \and D. Riccobelli$^{1,\,2}$ \and S. Gazzoni$^{1}$ \and  P. Vitullo$^{1}$ \and F. Ballarin$^{3}$ \and J. Falco$^{4}$ \and F. Acerbi$^{4}$ \and A. Manzoni$^{1}$ \and  P. Zunino$^{1}$ \and P. Ciarletta $^{1}$
}

\date{\footnotesize $^1$MOX -- Dipartimento di Matematica, Politecnico di Milano, Piazza Leonardo da Vinci 32, 20133, Milano, Italy\\
  $^2$Mathematics Area, mathLab, SISSA, Via Bonomea 265, Trieste, 34136, Italy\\
  $^3$Dipartimento di Matematica e Fisica ``N. Tartaglia'', Università Cattolica del Sacro Cuore, Via Garzetta 48, Brescia, 25133, Italy\\
$^4$Department of Neurosurgery – Fondazione I.R.C.C.S. Istituto Neurologico Carlo Besta, Via Celoria 11, Milano, 20133, Italy}

\renewcommand{\maketitlehookd}{%
  \begin{abstract}
    \noindent Glioblastoma is among the most aggressive brain tumors in adults, characterized by patient-specific invasion patterns driven by the underlying brain microstructure. In this work, we present a proof-of-concept for a  mathematical model of GBL growth, enabling real-time prediction and patient-specific parameter identification from longitudinal neuroimaging data.

    \noindent The framework exploits a diffuse-interface mathematical model to describe the tumor evolution and a reduced-order modeling strategy, relying on proper orthogonal decomposition, trained on synthetic data derived from patient-specific brain anatomies reconstructed from magnetic resonance imaging and diffusion tensor imaging. A neural network surrogate learns the inverse mapping from tumor evolution to model parameters, achieving  significant computational speed-up while preserving high accuracy.

    \noindent To ensure robustness and interpretability, we perform both global and local sensitivity analyses, identifying the key biophysical parameters governing tumor dynamics and assessing the stability of the inverse problem solution. These results establish a methodological foundation for future clinical deployment of patient-specific digital twins in neuro-oncology.
  \end{abstract}
}

\AtEveryBibitem{\clearfield{month}}
\AtEveryCitekey{\clearfield{month}}
\usepackage{amsthm, amsmath,dblfloatfix,siunitx}

\graphicspath{{figures}}

\newcommand{\vect}[1]{\mathbf{#1}}
\begin{document}

\maketitle %

\section{Introduction}
Glioblastoma (GBL) is one of the  deadliest types of brain cancer in adults \cite{Wen_2020}. Peculiar histological features consist of prominent cellular and nuclear atypia, numerous mitotic figures, necrosis, and microvascular proliferation \cite{smith2022major}. The 2021 WHO classification of central nervous system (CNS) tumors updated the diagnostic criteria introducing relevant innovations in tumor's definition, underlining the preponderant role of genetic elements compared to morphological ones \cite{Louis_2021}. GBL diagnosis is appropriate for each IDH wild-type astrocytic tumor presenting with concurrent +7/-10, EGFR amplification, or a TERT promoter mutation, even in the absence of classical high-grade histopathologic features  \cite{whitfield2022classification,weller2021eano}. As the most aggressive malignant glioma, GBL has a high invasive potential and grows along white matter fibres or vessels, imitating the physical structures of the brain extracellular environment \cite{Deisboeck_2001}. The most relevant consequence of this capability of extensive infiltration is that, despite aggressive multimodal therapy consistent in surgical resection, radiotherapy, and chemotherapy (Stupp protocol, as described in 2005), GBL invariably recurs, usually growing at the margin of the surgical cavity \cite{Stupp_2005}; accordingly, prognosis remains poor with a median progression-free and overall survival times approximately of 7 and 15 months, respectively, and the five-year survival rate is approximately 5\% \cite{Omuro_2013}.
Maximal tumor safe resection is the therapeutic cornerstone \cite{Brown_2016}: intraoperatively, GBL appears as an infiltrative mass, poorly delineated, bleeding, of increased consistency with peripheral grayish aspect and a central area of yellowish necrosis due to myelin breakdown. The ambiguous delimitation of tumor margins is one of the causes of difficult and rarely occurring complete tumor resection \cite{Rivera_2021}. The recent development of technological tools such as neuronavigation, optical fluorescence imaging, intraoperative brain magnetic resonance imaging (MRI), confocal laser endomicroscopy, and ultrasound have improved the intraoperative guidance, even they seem to be not enough to guarantee a radical resection \cite{Cavallo_2018,Hervey_Jumper_2016}. Radiologically, GBL appears as a bulky mass with heterogeneous enhancement and central necrosis; concomitantly surrounding T2/FLAIR abnormality indicates areas of vasogenic edematous, infiltrating, and non-enhancing neoplastic tissue: more than 90\% of tumor recurrences will occur within this T2/FLAIR envelope and there is limited research focused on the assessment of this region and its microenvironment. The development of MR spectroscopy allowed to measure the levels of specific brain metabolites which correlate with neoplastic aggressive proliferation. Positron emission tomography (PET) with specific traced amino acids can further improve the radiological diagnostic accuracy, increasing the visualization of highly metabolically active tissue and differentiating these regions from edematous tissue and post-treatment tissue \cite{Petrecca_2012}. Artificial Intelligence (AI) has recently emerged as a promising tool to further improve the neuroradiological power of tumor detection: in particular, radiomics treats images as numerical data and extracts intricate features, eluding human observation; AI’s impact could consequently extend to treatment planning for clinicians \cite{Chirica_2023}. The only way forward is interdisciplinary collaboration to define the best decision-making algorithms.
As stated above, the complex intratumoral heterogeneity at the genetic, biological, and functional levels, together with the tumor microenvironment is a crucial factor in making GBL extremely resistant to treatments \cite{Lem_e_2015,Chirizzi_2023}. In addition, GBL cells show a bursting tendency to infiltrate into the surrounding normal brain tissues of the tumor with a high complexity in tailoring surgical resection and adjuvant therapies \cite{Marko_2014}.
Mathematical modeling has played a crucial role in understanding GBL progression and treatment response, leveraging various approaches such as ordinary differential equations, partial differential equations (PDEs), agent-based models, and quantitative systems pharmacology frameworks. Classical models have been employed to predict patient-specific tumor dynamics and optimize therapeutic strategies. For instance, Swanson et al. \cite{swanson2008mathematical} developed a diffusion-reaction model to estimate tumor proliferation and infiltration from neuroimaging data, providing a foundation for predictive oncology. More recently, Gallaher et al. \cite{gallaher2020from} explored how cellular heterogeneity influences GBL growth and therapy response, emphasizing the impact of microenvironmental factors. Computational studies have also integrated immune system dynamics, as demonstrated by Storey et al. \cite{storey2020modeling}, who proposed a model combining oncolytic viral therapy and immune checkpoint inhibitors. Furthermore, spatial agent-based models, such as the one presented by Surendran et al. \cite{surendran2023agent}, have revealed key interactions between tumor cells and immune components under temozolomide and checkpoint blockade treatments. Additionally, Mongeon et al. \cite{mongeon2024spatial} highlighted the role of the tumor microenvironment in shaping immunotherapy outcomes through advanced spatial computational models. In this context, our approach builds upon these existing frameworks by integrating a diffuse-interface PDE model with machine learning techniques to enable rapid patient-specific parameter estimation and predictive tumor evolution modeling. This hybrid methodology bridges the gap between mechanistic modeling and data-driven learning, providing a novel computational tool for clinical decision-making.

In this work, we propose a computational framework for patient-specific prediction GBL growth, leveraging a reduced-order mathematical model informed by neuroimaging data. While the model does not explicitly incorporate therapeutic interventions, this choice is intentional, as our focus is on constructing the surrogate neural network model. However, the approach is directly applicable to integrating treatment effects in future developments, making it a flexible tool for clinical decision-making. \cite{Falco_2021}.

The remainder of this article is organized as follows. In Section 2, we introduce the mathematical model for GBL growth based on a diffuse-interface approach, followed by the finite element formulation and the reduced order model using Proper Orthogonal Decomposition. Section 3 presents the results of the numerical simulations, including patient-specific parameter estimation through neural networks. In Section 4, we discuss a proof-of-concept application on clinical data, demonstrating the potential of our framework. Finally, Section 5 concludes the paper with a summary of the key findings and directions for future research.

\section{Mathematical and numerical modeling of tumor growth}
In this section, we introduce a diffuse interface model and a numerical framework designed for predicting the patient-specific GBL growth from neuroimaging data.
Although certain information, such as the geometry of the patient's brain and local brain fiber orientation and physiological brain data, can be derived from imaging techniques as discussed in the Appendix~\ref{appendix_DT}, the cancer phenotype exhibits considerable variability and the prediction of its evolution is difficult and requires a patient-specific approach. Hence, we put forward a mathematical model alongside a machine learning-based approach to estimate the parameters pertaining to the patient's tumour evolution.

\subsection{Diffuse interface model of GBL growth}

In our model, we adopt a \textit{diffuse interface} approach to describe GBL growth within brain tissue based on mixture theory \cite{Bowen_1976}, following \cite{Anderson1998,Wise2008, Colombo2015}.
In this context, the brain is modeled as a mixture of two different constituents, also called \emph{phases}. One phase corresponds to the \textit{tumor cellular component} and the other phase represents the \textit{host brain tissue}, including healthy parenchymal cells and the extracellular matrix.  Let $\Omega\subset\mathbb{R}^3$ be the domain that represents the brain.
We can define the spatial concentration of each constituent at the instant of time $t$ at each point $\vect{x}\in\Omega$ \cite{ateshian2007theory}. The two phases are characterized by their respective volume fractions, $\phi_c(\mathbf{x},t)$ for the tumor and $\phi_l(\mathbf{x},t)$ for the healthy tissue, such that $\phi_c(\vect{x},t),\phi_l(\vect{x},t) \in [0,1]$, constrained by the saturation condition $\phi_c + \phi_l = 1$ at every point $\mathbf{x}$ in the domain at any time $t \in [0,T]$. This formulation enables the representation of tumor invasion and proliferation within a continuously deformable environment, where the interface between tumor and healthy tissue is diffused rather than sharply delineated.
The host brain tissue phase accounts for all non-tumoral components, including normal parenchymal cells, vascular structures, and extracellular matrix, which provide a deformable medium into which tumor cells infiltrate. Unlike solid tumor models that explicitly track multiple cell types (e.g., immune cells, stroma), our model focuses on the competition between proliferating tumor cells and the surrounding host tissue without explicitly resolving additional cellular populations. However, the interaction with the microenvironment is implicitly embedded in the model through parameters governing tumor proliferation and migration. This mathematical formulation allows us to capture the invasive nature of GBL growth while maintaining computational efficiency, which is crucial for patient-specific modeling and real-time clinical applications.

Using the saturation condition, it is possible to define a new phase-field variable $\phi:=\phi_c-\phi_l$ that assumes value $1$ in the tumour area and $-1$ on healthy areas. Since the brain is mainly composed of water, around 75-80\% in both healthy and tumor regions, we can reasonably assume that the two phases have a density roughly equal to the one of water $\gamma$.
Let $\vect{v}_c$ and $\vect{v}_l$ be the velocities of the cellular and the healthy phase, respectively. The following form of the mass balance holds true
\begin{equation}
  \label{eq:cont_c_l}
  \frac{\partial\phi_i}{\partial t} + \nabla \cdot (\phi_i\vect{v}_i)=\frac{\Gamma_i}{\gamma}  \quad  i \in \{c,l\},
\end{equation}
where $\Gamma_c$ and $\Gamma_l$ denote the mass source/sink terms per unit volume of the two fractions. Specifically, $\Gamma_c$ accounts for the proliferation of tumor cells, as well as their loss due to necrosis or apoptosis, while $\Gamma_l$ describes the corresponding changes in the healthy brain tissue.
To enforce the incompressibility of the whole mixture, we prescribe that $\Gamma_c=-\Gamma_l$.
Indeed, we can introduce the average velocity of the mixture as $\vect{v} = \phi_c\vect{v}_c + \phi_l\vect{v}_l$. If we sum the two continuity equations in \eqref{eq:cont_c_l}, we obtain
\[
  \nabla \cdot (\phi_c\vect{v}_c + \phi_l\vect{v}_l)=\nabla \cdot \vect{v}=0.
\]
By subtracting the two continuity equations  \eqref{eq:cont_c_l} we get
\begin{equation}
  \label{eq:cont_sub}
  \frac{\partial(\phi_c-\phi_l)}{\partial t} + \nabla \cdot (\phi_c\vect{v}_c -\phi_l\vect{v}_l)=\frac{\Gamma_c-\Gamma_l}{\gamma}
\end{equation}
Let $\mathcal{J}_c$ and $\mathcal{J}_l$ be  the mass fluxes of the two phases with
respect to the mixture velocity $\vect{v}$, defined as
\begin{align}
  \label{eq:Jc}
  \mathcal{J}_c&=\gamma \phi_c (\vect{v}_c-\vect{v}),\\
  \label{eq:Jl}
  \mathcal{J}_l&=\gamma \phi_l (\vect{v}_l-\vect{v}).
\end{align}
By introducing $\mathcal{J} = \frac{1}{\gamma}(\mathcal{J}_c-\mathcal{J}_l)$ and subtracting Eqs.~\eqref{eq:Jc}-\eqref{eq:Jl}, we get
\begin{equation}
  \label{eq:sopra}
  \phi_c \vect{v}_c-\phi_l \vect{v}_l= \phi \vect{v} + \mathcal{J}.
\end{equation}
We can use Eq.~\eqref{eq:sopra} to rewrite Eq.~\eqref{eq:cont_sub} as follows
\[
  \frac{\partial \phi}{\partial t}+\nabla \cdot (\phi \vect{v}) +\nabla \cdot \mathcal{J} = \frac{\Gamma}{\gamma} \quad \text{with} \quad \Gamma=\Gamma_c-\Gamma_l.
\]

We assume that the mixture is very viscous and free of external forces \cite{byrne2003modelling}. We use an approach based on non-equilibrium thermodynamics to determine a constitutive law for the mass fluxes.
We take the following expression of the Landau free energy:
\begin{equation}
  \label{eq:F}
  F(\phi)=\int_{\mathcal{B}_t}\Big( \kappa\Psi(\phi) + \frac{\epsilon^2}{2}|\nabla\phi|^2\Big) \,d\mathcal{B}_t,
\end{equation}
$\kappa$ is the brain Young modulus, $\epsilon$ defines  the interfacial tension, and  $\mathcal{B}_t$, i.e. the region occupied by the brain, is assumed to be with fixed boundaries over time. Therefore, from now on we omit the time specification and we refer to the domain with the symbol $\mathcal{B}$.
The two addends inside the integral in Eq.~\eqref{eq:F} represent the mixing energy density and the interface energy arising from the interaction between the two different phases, respectively \cite{ZAMM18}.

In this specific case, we take as cell-cell interaction
potential $\Psi(\phi)$ a function with a double-well shape, such that its minima are attained in $\phi=1$ and $\phi=-1$, corresponding to the two pure phases. A simple admissible choice is given by
\begin{equation}
  \label{interaction_potential}
  \Psi(\phi)=\frac{1}{4}(1-\phi^2)^2.
\end{equation}
Such a constitutive choice dictates that the system will evolve towards two stable equilibrium states, $\phi=\pm 1$, corresponding to the pure states of tumor and healthy tissues.

By following Fick's law, we postulate $\mathcal{J}$ to be proportional to the gradient of a chemical potential $\mu=\frac{\delta F(\phi)}{\delta \phi}$ , where $\delta$ is the Gâteaux functional derivative \cite{pozzi22}.
Thus, we assume that
\[
  \mathcal{J}=-\frac{1}{M_0}\mathsf{T}\nabla\mu.
\]
Here, $M_0$ is a friction coefficient, representing the resistance that the tumor experiences as it invades the host tissue: a higher $M_0$ corresponds to reduced mobility of tumor cells, while lower values facilitate their infiltration. The tensor $\mathsf{T}$ represents the preferential motility tensor \cite{agosti2018personalized} and encodes the anisotropic properties of GBL migration, accounting for preferential movement along brain structures such as white matter tracts and perivascular spaces.

To close the model, we prescribe a functional form for $\Gamma$. Tumor growth depends on the local oxygen concentration; in particular, cell duplication is promoted when oxygen is abundant, while apoptosis is enhanced under low oxygen availability. A suitable choice is
\begin{equation}
  \Gamma=\Gamma(\phi,n)=\nu \gamma \Big(\frac{n}{n_s}-\delta\Big)h(\phi).
\end{equation}
Here, $\nu$ is the tumor cell proliferation rate, $n$ represents the local oxygen concentration, $n_s$ is a physiological reference value for oxygen in brain tissue, $\delta$ is the hypoxia threshold, and $h(\phi)$ is a function that regulates proliferation within the natural range of $\phi$. This assumption reflects the established dependence of GBL proliferation on oxygen availability: cell division is enhanced in well-oxygenated regions and suppressed under hypoxia. Modeling $\Gamma$ as oxygen-dependent captures this spatial regulation of tumor growth, consistent with such experimental observations \cite{Gorin2004Perinecrotic}.
In this work, we only focus on GBL growth, but the choice of the functional expression for $\Gamma$ can be easily adapted to account for the tumour response to standard adjuvant therapy \cite{agosti2018personalized} and to immunotherapy \cite{pozzi22}.

The function $h$ should be constitutively prescribed. It should suppress the proliferation of tumor cells when $\phi=-1 $, i.e. when the tumor is absent. A possible choice for $h$ is given by
\[
  h(\phi)=\max\left(\min\left(1,\frac{1}{2}(1+\phi)\right),0\right)%
\]
The function $h(\phi)$ is a continuous monotonically increasing function from 0 and 1, ensuring that proliferation is suppressed in healthy
tissue ($\phi = -1$), reaches its maximum in tumor regions ($\phi = 1$), and
transitions gradually in the intermediate region.

The evolution equations for the GBL and the nutrient concentrations are derived within a variational framework, where the mixture dynamics follow a gradient flow structure. In particular, the mass fluxes are postulated to be driven by the gradient of a chemical potential, arising from the variational derivative of a free energy functional, under the principle of maximal dissipation, as in \cite{garcke2016cahn,pozzi22}.

The dynamics of oxygen concentration  is governed by a reaction-diffusion equation, where $\mathsf{D}$ is the diffusivity tensor of the nutrient, $S_n$ is the \textit{oxygen supply rate} and $\delta_n$ is the \textit{oxygen consumption rate}.
Moreover, a Cahn-Hilliard type equation, where the mass exchanges  do not conserve the order parameter, describes the evolution of GBL growth. The resulting partial differential system reads:
\begin{equation}
  \label{eq:model}
  \left\{
    \begin{aligned}
      &\frac{\partial\phi}{\partial t} = \nabla \cdot \Big( \frac{1}{M_0}\mathsf{T}\nabla\mu\Big)+\nu\left(\hat{n}-\delta\right)h(\phi),\\
      &\mu = \kappa\Psi'(\phi)-\epsilon^2\Delta\phi, \\
      &\frac{\partial \hat{n}}{\partial t}=
      \begin{aligned}[t]
        &\frac{S_n}{3}\left(1-\hat{n}\right)(2-\phi)\\
        &+\nabla \cdot(\mathsf{D}\nabla \hat{n})-\delta_n \hat{n} h(\phi) .
      \end{aligned}
    \end{aligned}
    \right.
  \end{equation}
  where the auxiliary variable $\mu$ represents the chemical potential, while $\hat{n} = n/n_s$. The parameters are the tumor cells proliferation rate $\nu$, the tumor inter-phase friction $M_0$, the brain Young modulus $\kappa$, the diffuse interfacial energy $\epsilon$, the oxygen concentration in vessels $n_s$, the hypoxia threshold $\delta$, the oxygen consumption
  rate $\delta_n$ and the oxygen supply rate $S_n$. Their biological range is collected in Table \ref{tab:parameters}. The tensors $\mathsf{D}$ and $\mathsf{T}$ can be extracted from patient's imaging data following the procedure detailed in Appendix~\ref{appendix_DT}.
  Finally, we enforce a homogeneous Neumann boundary condition for each physical variable at the brain boundary to model no flux across the brain boundary, consistent with the typical behavior of primary GBL, which remains confined to the parenchyma and invades preferentially along white matter tracts. While this does not account for  events like metastatic spread or CSF-mediated transport, it is a reasonable simplification for the physiological regimes considered in this study.

  Henceforth, we refer to the task of finding the solution to Eq.~\eqref{eq:model} as the direct problem. For the sake of improving the computational efficacy, we propose in the following a neural network approach to construct a surrogate model based on model order reduction.

  \subsection{Full order model}

  \begin{table*}[t!]
    \centering
    \begin{tabular}{| c| c | c | c |}
      \hline
      \textbf{Symbol} & \textbf{Range of values} & \textbf{Ref.} \\
      \hline$M_0$  & $\SI{1.38e3}{}-\SI{5.03e3}{\pascal\day\per\milli\meter\squared}$ &\cite{swabb1974diffusion} \\
      \hline$\nu$  & $\SI{1.2e-2}{}-\SI{0.5}{\per\day}$ &  \cite{swanson2000quantitative, martinez2012hypoxic}  \\
      \hline$S_{n}$  & $\SI{1e3}-\SI{1e5}{\per\day}$ &\cite{chatelain2011emergence} \\
      \hline$\delta_{n}$  & $\SI{1e3}-\SI{1e5}{\per\day}$ &\cite{martinez2012hypoxic} \\
      \hline$\kappa$  & $\SI{1.06e2}{}- \SI{1.53e3}{\pascal}$ &\cite{clatz2005realistic} \\
      \hline$\delta$  & $0.1-0.33$ &\cite{bedogni2009hypoxia}  \\
      \hline
    \end{tabular}
    \caption{Biological range found in literature for the parameters of the model.}
    \label{tab:parameters}
  \end{table*}
  First, we computationally solve Eq.~\eqref{eq:model} using  the finite element method. In such a way, we obtain a discrete counterpart of the model proposed in Eq.~\eqref{eq:model}. We refer to such a discrete problem as the \emph{full order model} (FOM) on a discrete partition $\mathcal{T}_h$. Then, we divide the temporal interval $[0,T]$ into N discrete subintervals $\Delta t=T/N$. The $j$-th simulation time-point $t^j=j\Delta t$ with $j=0,...,N$. Next, we introduce the finite element space $V_h=\Bigl\{ \chi \in C^0(\Omega): \chi|_{K}\in \mathbb{P}^1(K) \quad \forall K \in \mathcal{T}_h \Bigr\} \subset H^1(\Omega)$, which is the space of continuous polynomial functions of degree 1 ($\mathbb{P}^1$) when restricted on each element $K$. $V_h$ is a subset of the Hilbert space $H^1(\Omega)$ that contains $L^2(\Omega)$ functions whose first weak derivative is in $L^2(\Omega)$ too.

  Thus, given the initial data $(\phi^0_h,\hat{n}^0_h) \in V_h \times V_h$ we obtain the following discrete problem:
  \begin{small}
  \[
    \left\{
      \begin{aligned}
        &
        \begin{aligned}
          \left(\frac{\phi_{h}^{j+1}-\phi_{h}^j}{\Delta t}, \varphi_{h}\right)= & -\frac{1}{M_{0}}\left(\mathsf{T}\nabla \mu_{h}^{j+1}, \nabla \varphi_{h}\right)+\\
          &+\nu\left(\left(\hat{n}_{h}^{j+1}-\delta\right) h\left(\phi_{h}^j\right), \varphi_{h}\right)
        \end{aligned} \\
        &
        \begin{aligned}
          \left(\mu_{h}^{j+1}, v_{h}\right) =  &\epsilon^2\left(\nabla \phi_{h}^{j+1}, \nabla v_{h}\right)+\kappa\left(\Psi_{c}^{\prime}\left(\phi_{h}^{j+1}\right), v_{h}\right)  +\\
          &+\kappa\left(\Psi_{e}^{\prime}\left(\phi_{h}^j\right), v_{h}\right)
        \end{aligned}\\
        &
        \begin{aligned}\left(\frac{\hat{n}_{h}^{j+1}-\hat{n}_{h}^j}{\Delta t}, q_{h}\right) =  &-\left(\mathsf{D}\nabla \hat{n}_{h}^{j+1}, \nabla q_{h}\right)+\\
          &+S_{n}\left(\left(1-\hat{n}_{h}^{j+1}\right) \frac{1}{3}\left(2-\phi_{h}^j\right), q_{h}\right)+\\
          & -\delta_{n}\left(\hat{n}_{h}^{j+1} h\left(\phi_{h}^j\right), q_{h}\right)
        \end{aligned}
      \end{aligned}
      \right.
    \]
    \end{small}
    where $(\cdot, \cdot)$ denotes the standard $L^{2}$ inner product over $\Omega$. As suggested in \cite{byrne2003modelling} we prescribe the following splitting for the Cahn-Hilliard potential to ensure the gradient stability of the scheme:
    \[
      \Psi_{c}\left(\phi_{h}^{j+1}\right)=\frac{\left(\phi_{h}^{j+1}\right)^{4}+1}{4},\quad\Psi_{e}\left(\phi_{h}^j\right)=-\frac{\left(\phi_{h}^j\right)^{2}}{2}.
    \]
    Decomposing the potential in such a way, i.e. in a convex term $\Psi_{c}$ that we can treat with an implicit scheme and a concave term $\Psi_{e}$ that is treated with an explicit scheme, ensures the solution to be stable over time \cite{tierra2015numerical}.

    \subsection{Reduced order model}\label{sec:reduced_order_model}
    Solving the FOM requires a huge amount of computational resources and time. Aiming at constructing an effective procedure to solve the inverse problem of patient-specific parameter identification from neuroimaging data,
    we resort to a \emph{reduced order model} (ROM) based on linear projections onto a lower dimensional subspace as a robust and more efficient solution strategy.
    The basic idea is to construct a reduced basis (RB) space for the approximation of the discrete solution manifold, that is, the set of all FOM solutions obtained for varying parameters within a given parameter space.
    Starting from the system Eq.~\eqref{eq:model}, we perform a Proper Orthogonal Decomposition (POD) \cite{HesthavenRozzaStamm2015,Quarteroni_2016} on a set of FOM solutions, named \textit{snapshots}.

    The construction of a basis for the final reduced order space consists of two similar steps.
    We first perform a Singular Value Decomposition (SVD) of the matrix collecting, columnwise, snapshots associated with the variable $f=\left\{\phi\,,\mu \,, n\right\}$ associated with a particular choice of the parameters $\mathcal{P}_k=[\nu_k, M_{0l}, \kappa_k, \delta_k, \delta_{nk}, S_{nk}]$ over time. Specifically, the matrix columns are the nodal values of the solution at a specific time-step
    \[
      F^1_f=[f_k^0,...,f_k^N],
    \]
    where $N+1$ is the number of time-steps.
    By applying SVD on $F^1_f$, we obtain a basis $\left\{ \xi_{kl}^f\right\}_{l=1,...,N_{\text{POD}}^{k}}$ from each set of parameters $\mathcal{P}_k$, where $N_{\text{POD}^k}$ is chosen such that information content that the POD basis should retain, indicated as $ic \in (0,1]$, is about $ic=0.95$ for each variable. We denote by $M$ the cardinality of the training set of selected parameters, so that $k=1,\,\dots,\,M$.
    Until this point, the bases contain most of the information on the evolution of the tumor through time for a specific set of parameters.
    Then, we perform another SVD, this time on the matrix collecting all the $M$ sets of basis functions obtained at the previous step, that is
    \begin{equation*}
      F^2_f=\left[\xi_{11}^f,...,\xi_{1N^1_{\text{POD}}}^f,...,\xi_{M1}^f,...,\xi_{MN^M_{\text{POD}}}^f \right].
    \end{equation*}
    The final result is a basis $\Big\{ \xi_l^f\Big\}_{l=1,...,N_\text{POD}}$ of the reduced order space for each variable $f=\left\{\phi\,,\mu \,, n\right\}$. A similar strategy has been used e.g.\ in \cite{Audouze2009,rapun2010reduced,himpe2018hierarchical,wang2019non} to generate the POD basis for models depending on both time and parameters.

    Denoting by $\xi_i^f$ the generic element of the reduced basis of the physical variable $f$, we can write
    \begin{equation*}
      \label{eq:proj}
      \ \phi_{h}^{t}=\sum_{i=1}^{N_{\text {POD }}} a_{\phi i}^{t} \xi_{i}^{\phi},\,  \mu_{h}^{t}=\sum_{i=1}^{N_{\text {POD }}} a_{\mu i}^{t} \xi_{i}^{\mu},\, \hat{n}_{h}^{t}=\sum_{i=1}^{N_{\text {POD }}} a_{n i}^{t} \xi_{i}^{n},
    \end{equation*}
    where $N_\text{POD}$ is the cardinality of the reduced basis.
    $N_{\text{POD}}$ is chosen such that information that the POD basis should cover $ic \in (0,1]$ is about $ic=0.95$ for each variable.

    To summarize, the steps that we perform for each phase are \cite{agosti2020learning}:
    \begin{itemize}
      \item prescribe the amount of required information that the POD basis should cover $ic \in (0,1]$;
      \item compute the trace $tr(F_f^tF_f)$ of the correlation matrix $F_f^tF_f=(f^m,f^l)_{ml}$;
      \item evaluate the pair eigenvalues-eigenvectors \[\{\lambda_{fi},\nu_f^i\}_{i=1,...,N_f^{\text{POD}}} \] of $F_f^tF_f$;
      \item $N_f^{\text{POD}}=\text{min}\left\{m,\left(\sum_{i\leq m}\lambda_i\right)/tr(F^tF)\leq ic\right\}$, that is the number of elements in the basis, is set;
      \item $N^\text{POD}=\text{max}\left\{N_\phi^{\text{POD}}, N_\mu^{\text{POD}}, N_n^{\text{POD}} \right\}$
      \item set $\xi_s^f = \frac{1}{\sqrt{\lambda_{fs}}}\sum_j (\nu_f^s)_jf^j $ where $(1\leq s\leq N^{\text{POD}})$.
    \end{itemize}

    Since the model in Eq.~\eqref{eq:model} is non-linear, a classical POD-Galerkin method would in principle require the projection of the non-linear operators. When a general non-linearity is present, the cost to evaluate the projected nonlinear function still depends on the dimension of the original system, resulting in simulation times that hardly improve over the original system. A possible approach to overcome this issue is to rely on suitable hyper-reduction techniques, such as those based on a greedy algorithm using DEIM interpolation, see e.g. \cite{Nonlinear,farhat2020} for further details. In this work, an alternative approach, exploiting neural networks, is preferred to approximate the RB coefficients in a non-intrusive framework, resorting only on the simulation data and without manipulating directly the governing equations with Galerkin projections as with the classical intrusive hyper-reduction techniques.

    The reduction of the problem to a few degrees of freedom, equal to the dimensionality of the reduced space \( N_\text{POD} \) and corresponding to the coefficients of the RB, makes it possible to train a simple neural network that maps the parameter space onto the space of the RB coefficients, a method usally referred to as the POD-NN approach \cite{Hesthaven}. Given a set of parameters  $\mathcal{P}=[\nu, M_0, \kappa, \delta, \delta_n, S_n]$ of cardinality \( N_P \), along with a temporal step \( t \), we train the neural network $\vect{NN}_\phi : \mathbb{R}^{N_\mathcal{P}+1}\rightarrow\mathbb{R}^{N_\text{POD}}$ to compute the coefficients \( \{a^{\phi}_{t,i}\} \in \mathbb{R}^{N_{POD}} \) for the RB of the tumor concentration variable \( \phi \). Following the procedure presented in \cite{Hesthaven}, $\vect{NN}_\phi$ is an approximation of the function that map points $[\nu, M_0, \kappa, \delta, \delta_n, S_n, t]$, which corresponds to a tumor distribution at a given instant \( t \), to the space of coefficients \( \{a^{\phi}_{t,i}\}_{i=1,...,N_{POD}} \) of the projected solution in the ROM space at the same time instant. We choose not to make \( \varepsilon \) vary since it is related to the thickness of the diffusive interface that is fixed a priori, while the tensors $\mathsf{T}$ and $\mathsf{D}$ are extracted from neuroimaging data, as described in the Appendix.

    \section{Results}

    \subsection{Surrogate approach to estimate patient-specific parameters from neuroimaging data}

\begin{figure*}[t!]
      \centering
      \includegraphics[width=0.97\textwidth]{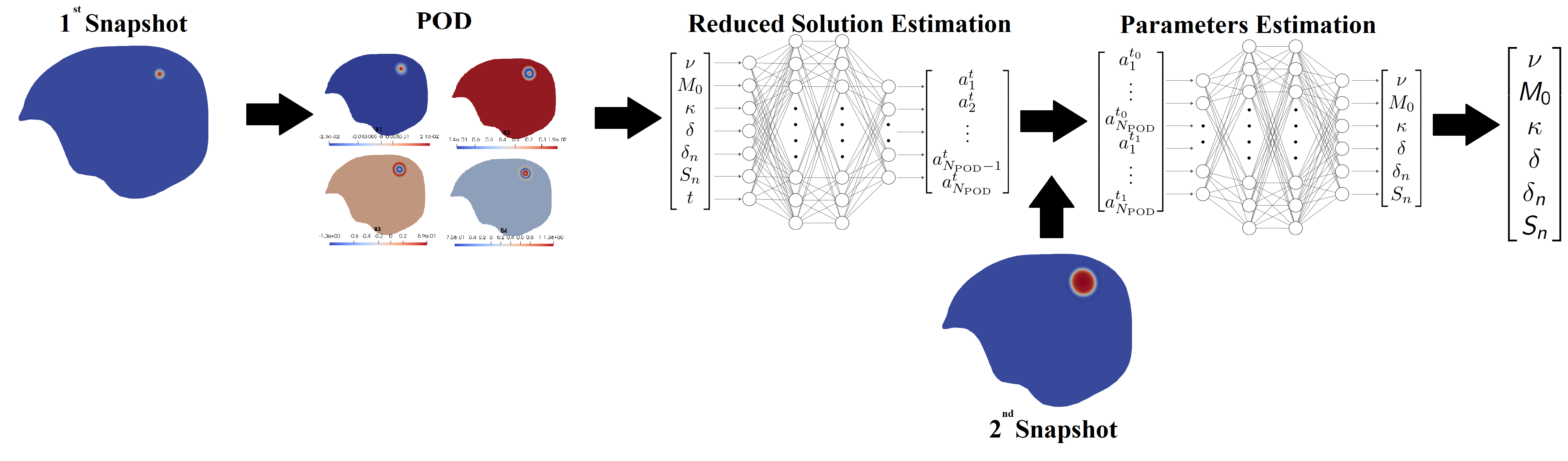}
      \caption{Representation of the computational pipeline. The geometry and the distribution of the tumor is known at for $t=t_0$. From this datum, we perform the POD and get the reduced order solution estimation for the direct problem (POD-NN procedure). Given the distribution of the grown tumor at $t=t_1=t_0+20$ days, we train a neural network to solve the inverse problem estimating the patient-specific parameters.}
      \label{fig:pipeline}
    \end{figure*}

    We propose in the following a numerical pipeline to infer the patient-specific parameters of GBL growth from the observed tumor distribution at two different instants of time, given by the clinical follow-up protocol summarized in the Appendix~\ref{appendix_protocol}.
    In the following, we refer to the identification of the patient-specific  parameters given the observed distributions of the tumor  as the \emph{inverse problem}. Also for this purpose, we exploit surrogate neural network techniques to approximate the solutions. An illustration of the proposed computational pipeline is presented in Fig.~\ref{fig:pipeline}

    \noindent In order to solve the inverse problem of estimating the
    patient-specific parameters from the observed tumor distribution given by two instants of time from clinical follow-up, we construct a second neural network.
    In this case, the trained neural network is a map $\vect{NN}_\text{inv}: \mathbb{R}^{2 N^\text{POD}}\rightarrow\mathbb{R}^{N_\mathcal{P}}$ that takes as inputs  two tumor distributions, identified with their projection coefficients over the reduced basis, and gives the set of parameters as the output, i.e.
    \begin{small}
      $$
      \left(\nu, M_0, \kappa, \delta, \delta_n, S_n\right)= \vect{NN}_\text{inv} \left( a_{\phi 1}^{t_0},...,a_{\phi N_\text{POD}}^{t_0},a_{\phi 1}^{t_1},...,a_{\phi N_\text{POD}}^{t_1}\right)
      $$
    \end{small}
    where $t_0$ and $t_1=t_0+(20\,\text{days})$ represent the time interval that elapses from the first and second MRI.\\
    \noindent We choose the weighted sum as the propagation function, the LeakyReLU as the activation function and just the identity for the output function.
    Moreover, the loss function used is the mean squared error.

    \subsection{Proof-of-concept}
    We finally build a proof-of-concept by image segmentation of the MRI and DTI data from a clinical case,  as shown in Fig.~\ref{fig:mesh_ref}. The details of the clinical and radiological protocols are summarized in the Appendix~\ref{appendix_protocol}. This realistic brain-shaped mesh has 48434 vertices and 280399 tetrahedral elements. A mesh refinement is performed in the neighborhood of the initial placement of the tumor.
    For each simulation, a piecewise linear basis function is chosen, so that the degrees of freedom of the solution correspond to the number of vertices.
    For the numerical solution of the FOM, we rely on a HPC cluster (Intel\textsuperscript{®} Xeon\textsuperscript{®} Processor E5-2640 v4, 20 cores, 64 GB RAM).   The overall implementation framework exploits the functionalities given by the platform \texttt{FEniCSx}, a popular open-source environment for solving partial differential equations.

    \begin{figure}[b!]
      \centering
      \subfloat[]{\includegraphics[width=0.46\columnwidth]{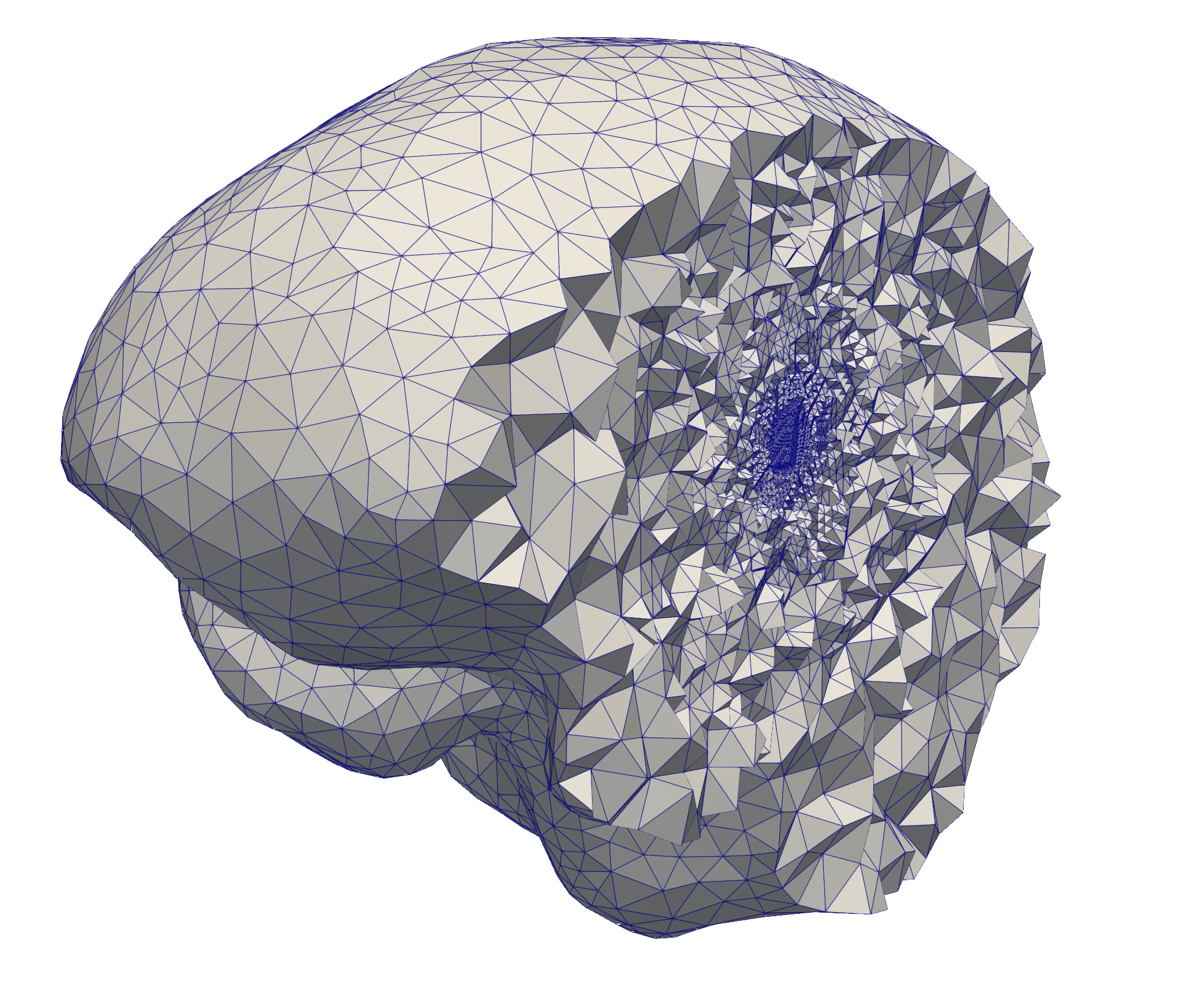}}
      \subfloat[]{\includegraphics[width=0.46\columnwidth]{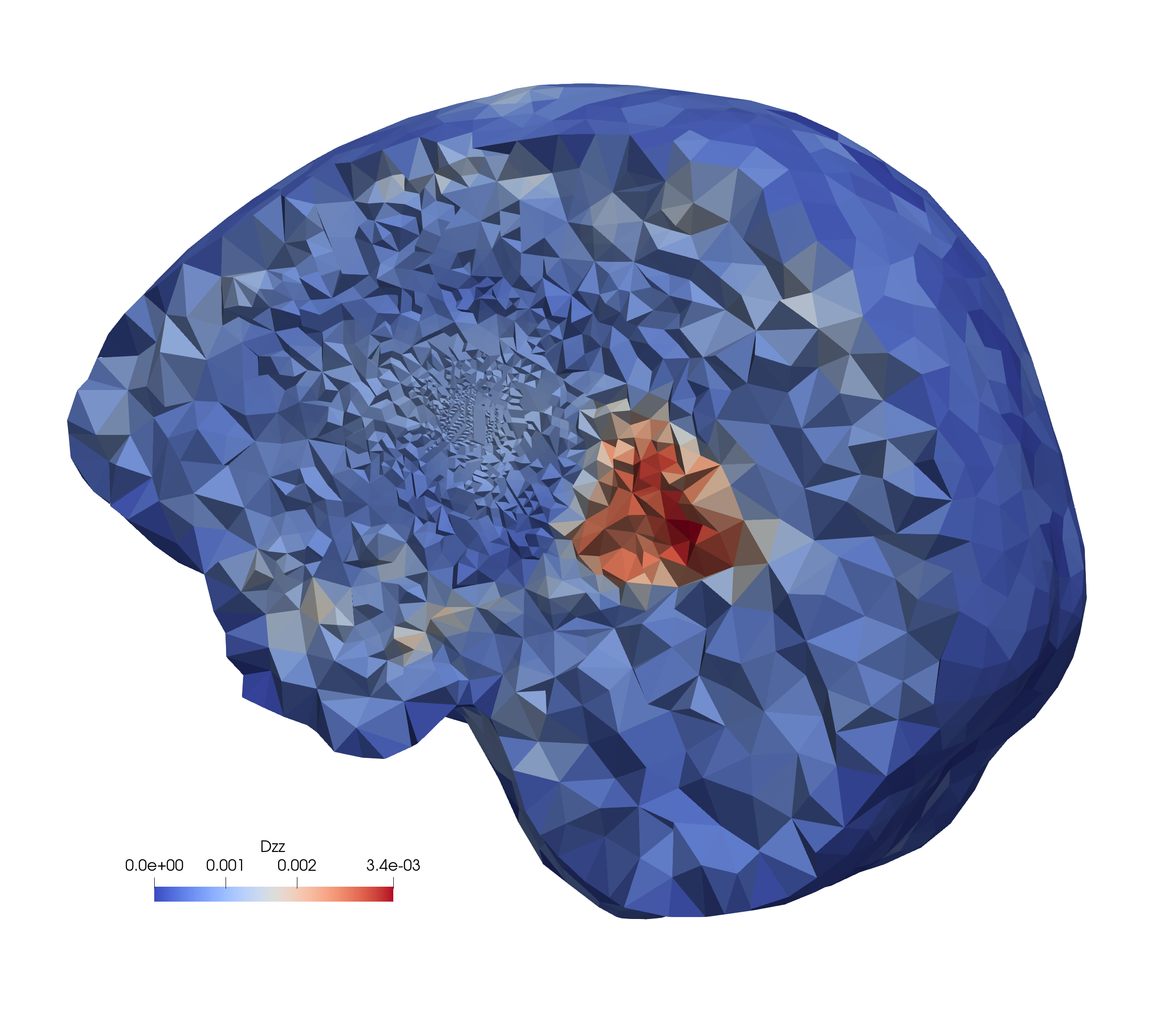}}
      \newline
      \subfloat[]{\includegraphics[width=0.35\columnwidth]{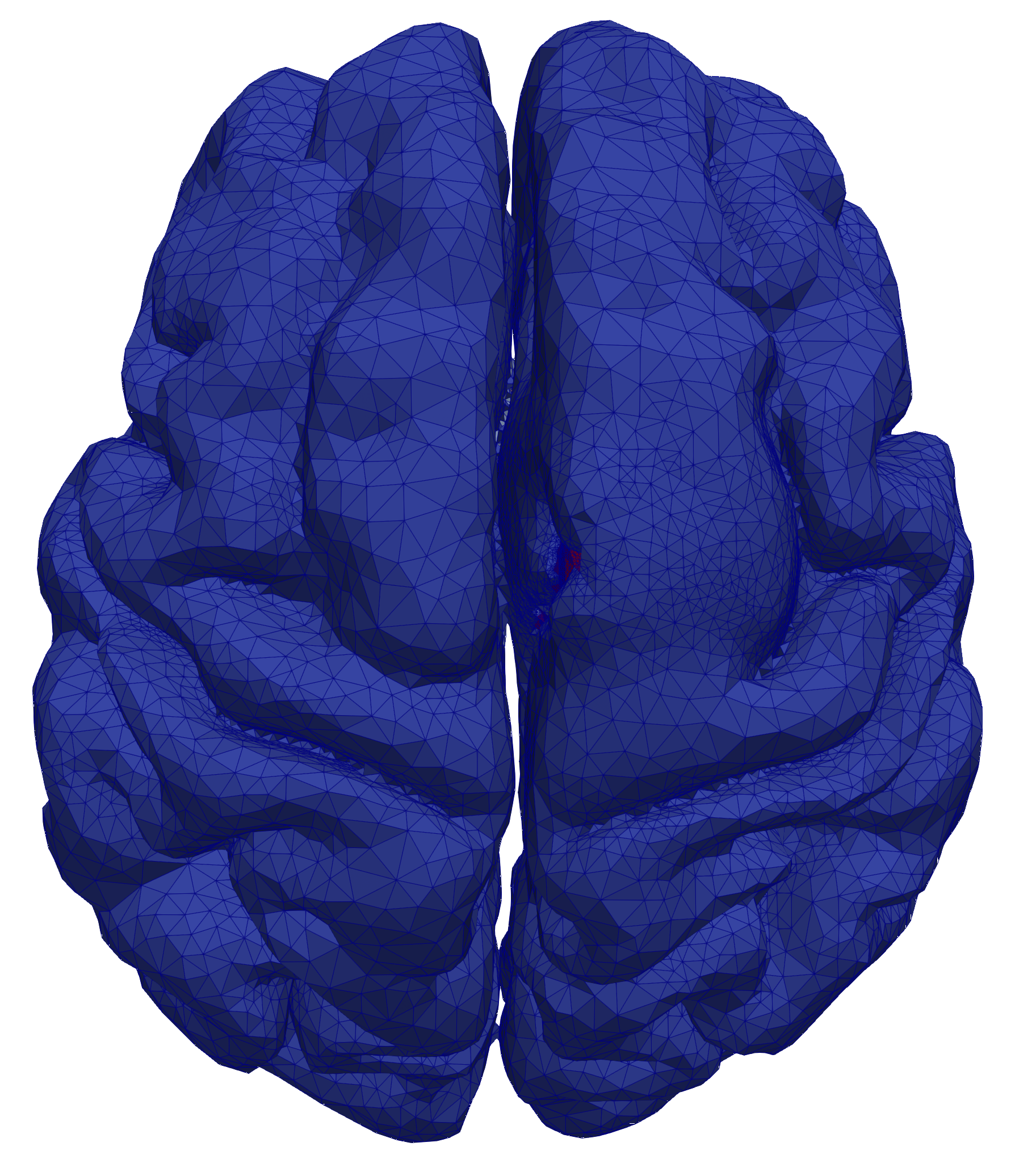}}
      \subfloat[]{\includegraphics[width=0.46\columnwidth]
      {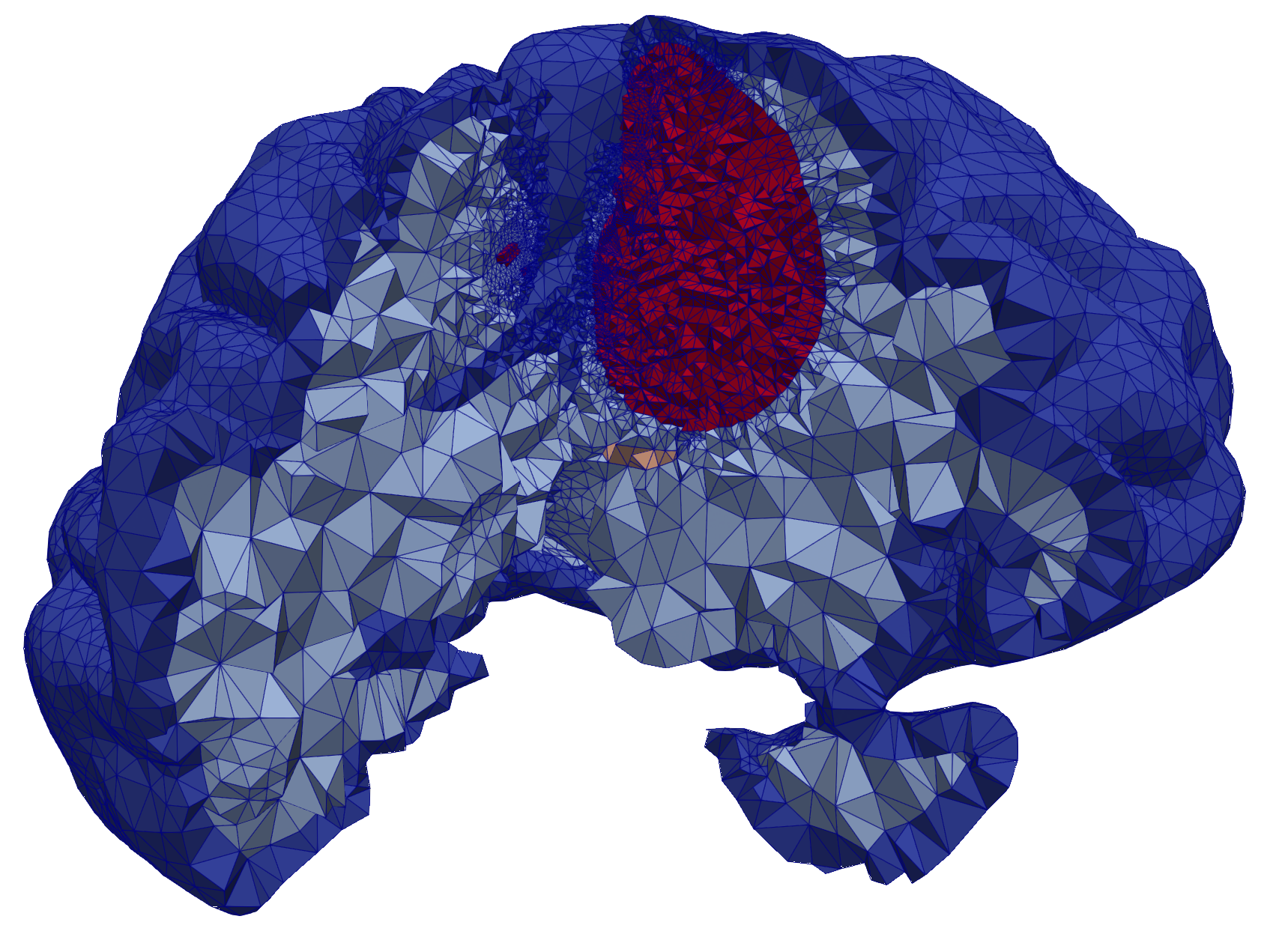}}
      \caption{ A representation of the computational domain extracted from the  MRI data (a) with superposed color map of a component of the tensor $\mathsf{D}$ extracted from DTI data in \SI{}{\milli\meter\squared\per\day} (b);  reconstructed domain of the brain cortex (c)  and corresponding  sections  for grey matter (blue), white matter (light blue), and tumour (red) in the labeled mesh (d).}
      \label{fig:mesh_ref}
    \end{figure}

    The implementation of the used code heavily relies on two of its components: \texttt{dolfinx}, a C++/Python library providing data structures and algorithms for finite element meshes, automated finite element assembly, and numerical linear algebra, and the Unified Form Language \texttt{UFL} which is a domain-specific language for declaration of finite element discretization of variational forms.
    The construction of the ROM basis is obtained through  \texttt{RBniCSx}, a library useful to implement reduced order modeling techniques. The neural network is implemented in Python using \texttt{PyTorch}. As a minimization procedure for the loss function we have used the L-BFGS algorithm \cite{LiuNocedal}.

    For training the neural network of the direct problem, we draw parameters out of the biological range exhibited in Tab.~\ref{tab:parameters}.

    To obtain adequate accuracy the training for the direct POD-NN, we construct a data set from numerical simulations obtained by 750 different sets of parameters. Using 60 temporal steps, each of them representing 0.5 days, we finally get $N_\text{Data}^\text{dir}=45000$ input-output pairs.  This data set is split into a training set with $N^\text{dir}_\text{train}=33000$ elements and a test set with $N^\text{dir}_\text{test}=12000$ elements.

\begin{figure*}
      \centering
      \includegraphics[width=0.33\textwidth]{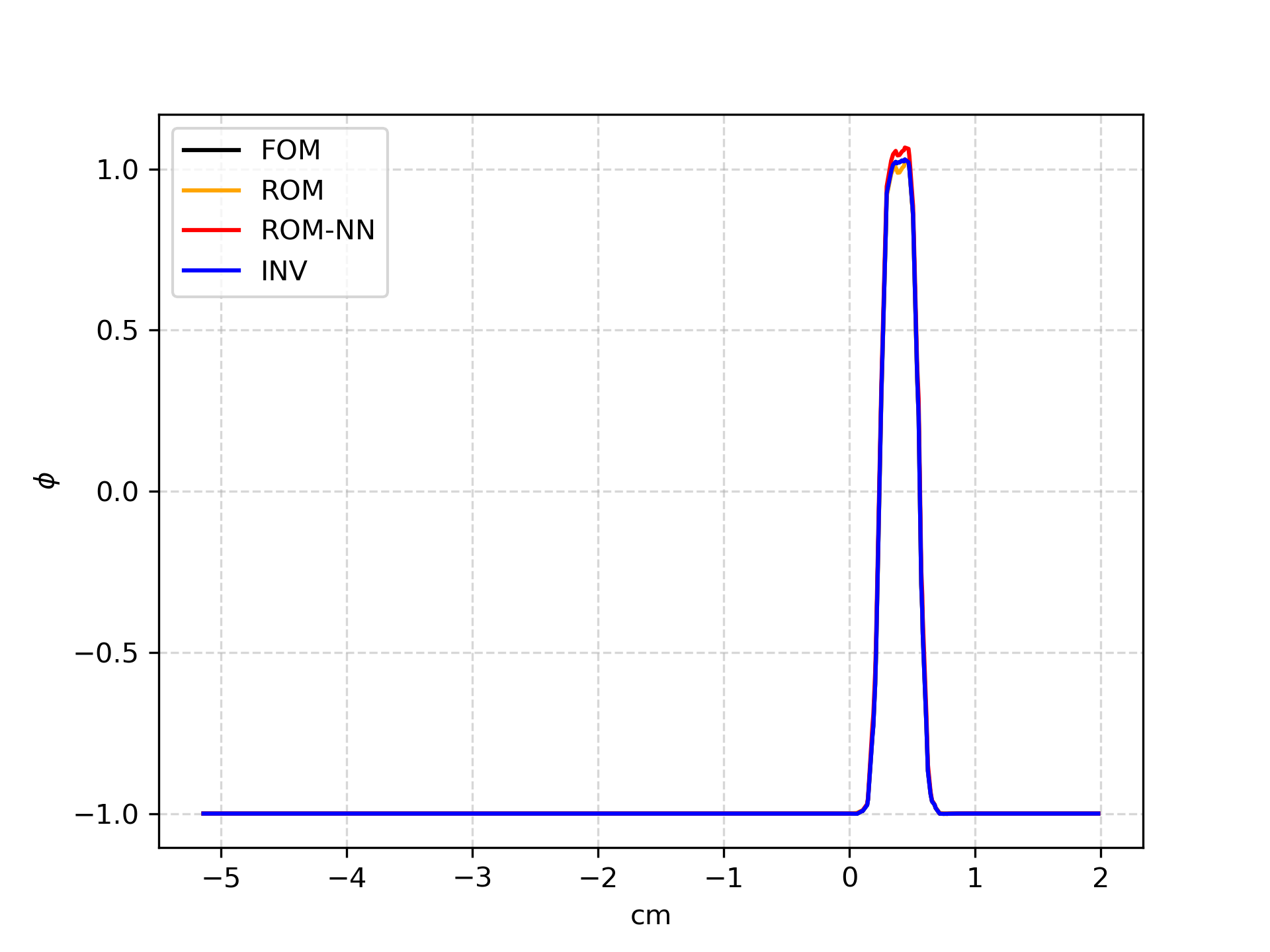}\hfill
      \includegraphics[width=0.33\textwidth]{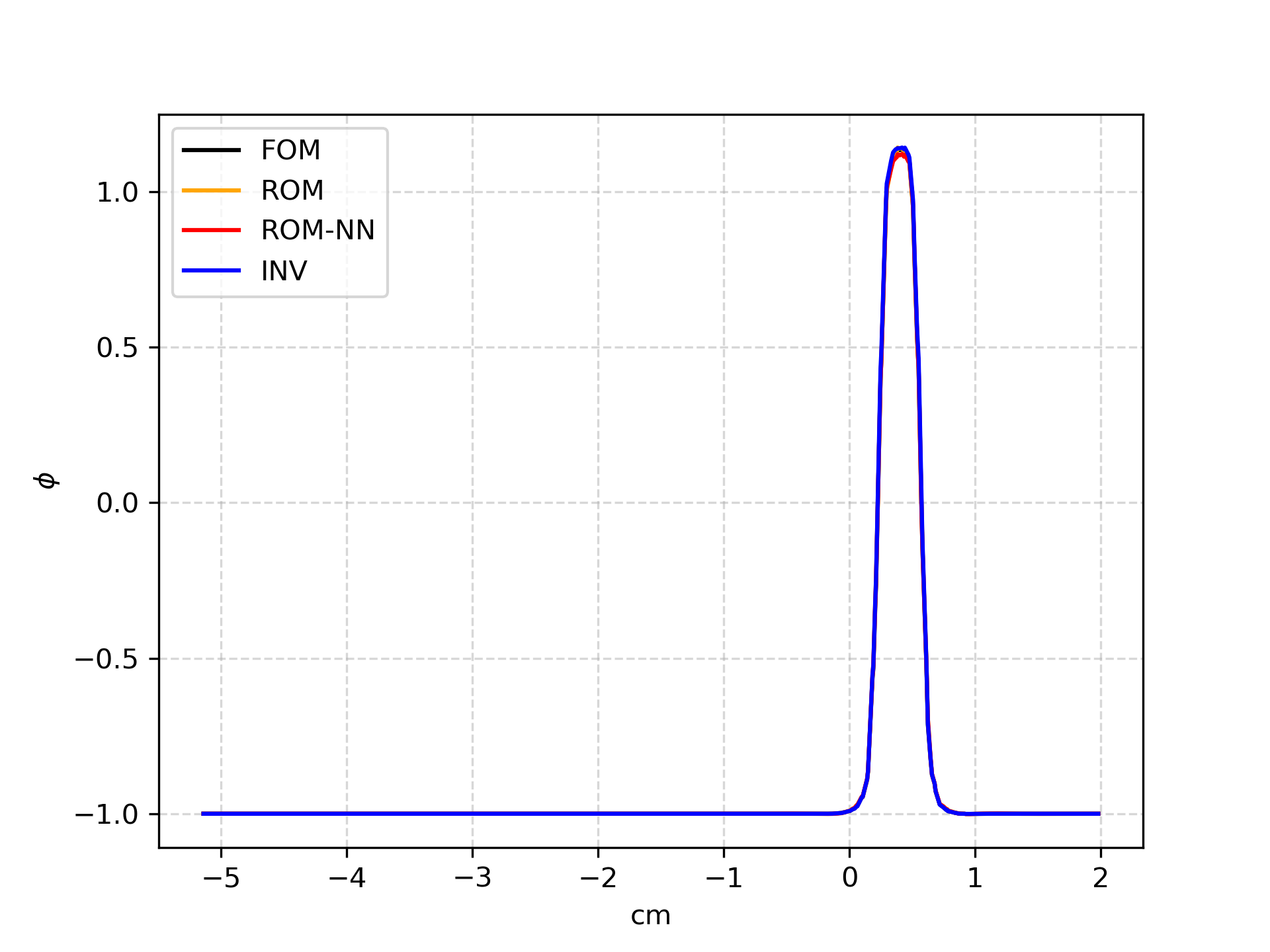}\hfill
      \includegraphics[width=0.33\textwidth]{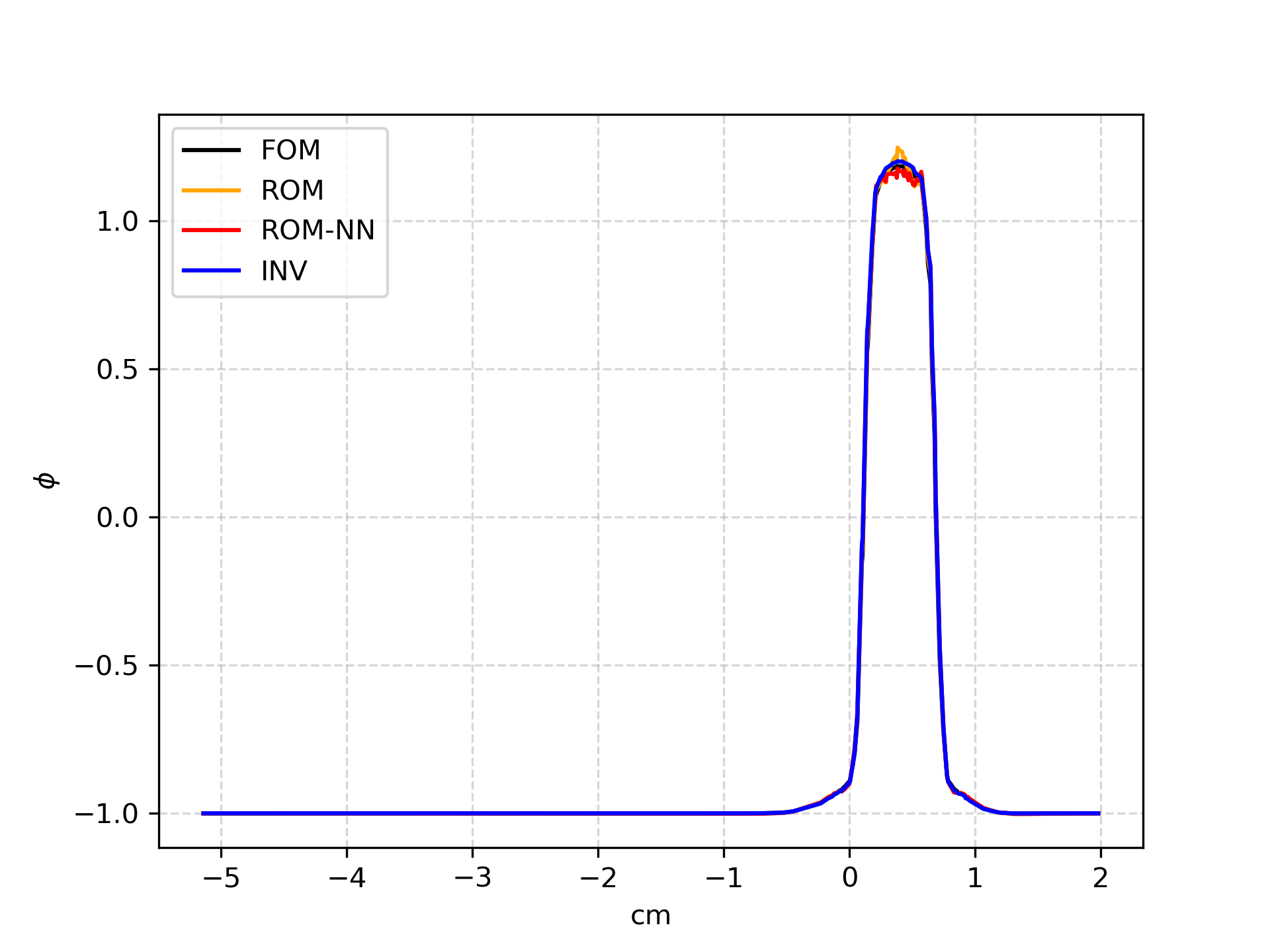}\hfill
      \caption{ Plot of the solution $\phi$ within a fixed sagittal plane intersecting the tumor centroid at t=\,0 (left),\,15 (center),\,30  days (right).  Solid lines indicate the FOM solution (black), the POD-Galerkin solution (orange), the  POD-NN solution (red), and the FOM solution obtained using  the parameter obtained in the inverse problem (blue). }
      \label{fig:comper_plot}
    \end{figure*}

    We perform FOM computations with $M=64$ different sets of parameters, to build up a
    representative  basis that can retain most of the energy present in all of the original variables.
    In this case, a basis with $N_\text{POD}=20$ elements was big enough to have an acceptable error between the FOM solution and the POD-Galerkin one, as shown in Fig.~\ref{fig:comper_plot}.

\begin{figure}[t!]
      \centering
      {%
      \includegraphics[width=\columnwidth]{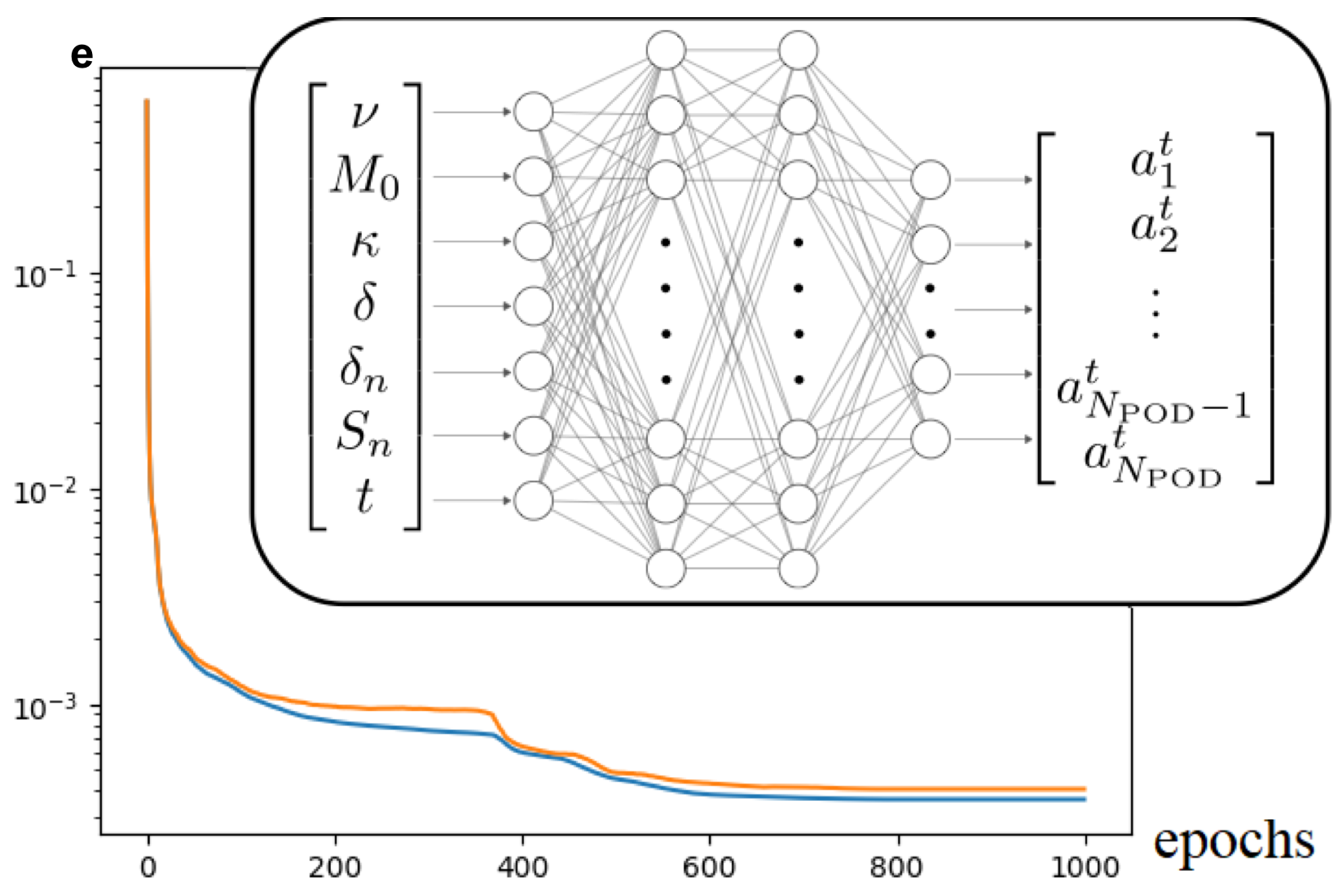}}
      \hspace{0.5em}
      \includegraphics[width=\columnwidth]{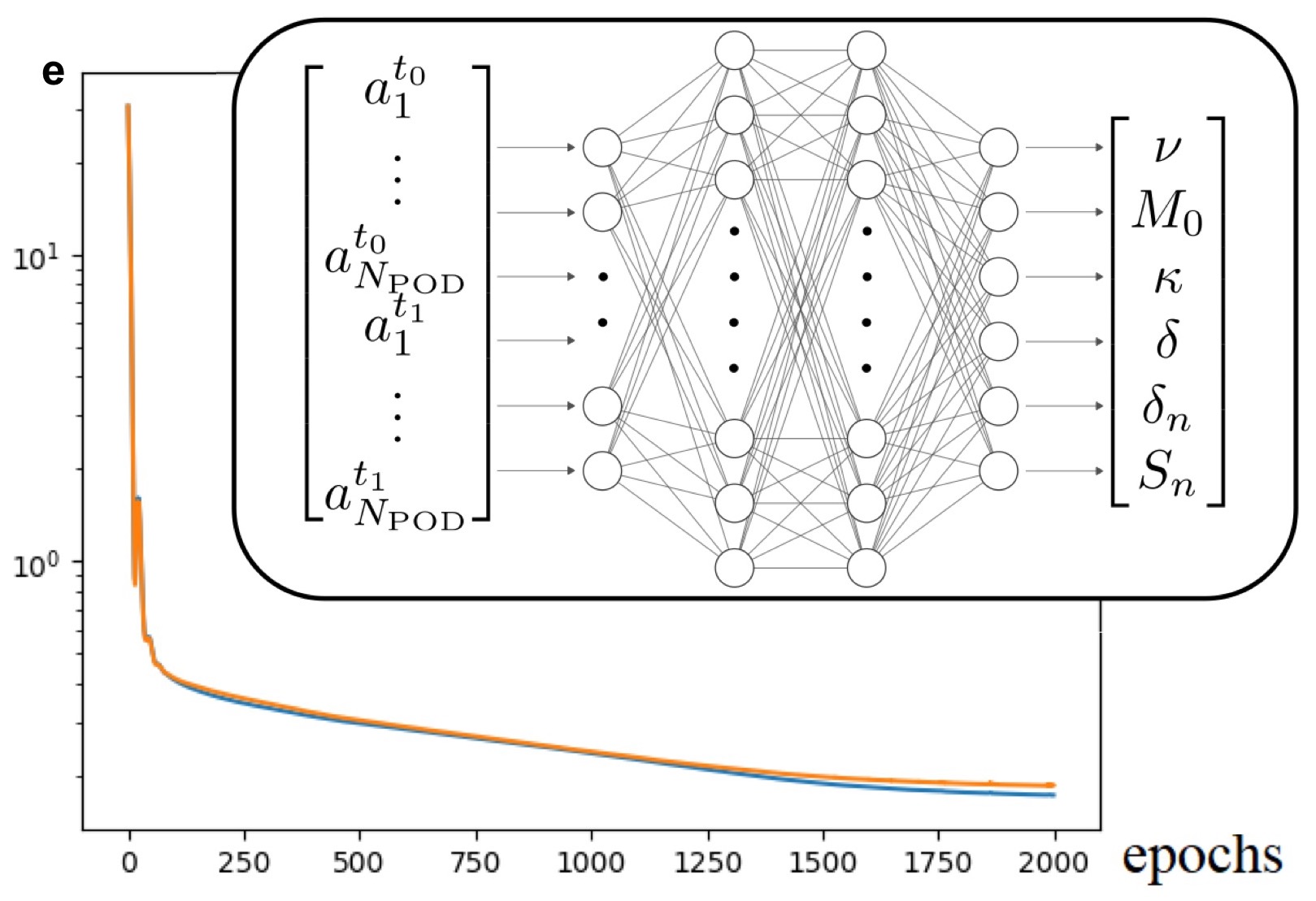}
      \caption{Absolute (top) and relative (bottom) mean squared error $\mathbf{e}$ over the epochs in the training of the direct (top) and the inverse (bottom) neural networks. The solid lines indicate the errors over the training set (blue) and over the test sets (orange).}
      \label{fig:err_inv}
    \end{figure}

    From this, it is possible to create a data set to train the neural networks $\vect{NN}_\phi$ surrogating the map of the direct problem. Training data refer to 750 different possible evolutions of the tumor starting from the same initial condition $\phi_0(x,y,z)= 2 e^{-100({(x-25)^2+(y-4)^2+(z-30)^2)}^2}-1$ where spatial quantities are measured in $mm$. The results of the training in terms of mean squared error over epochs are shown in Fig.~\ref{fig:err_inv}.

\begin{figure*} [ht!]
      \centering
      \begin{center}
        \begin{tabular}{ c c c c }
          & \textbf{t=0 d}& \textbf{t=15 d} & \textbf{t=30 d} \\
          \hline
          & & Full Order Model & \\
          \includegraphics[width=0.07\textwidth]{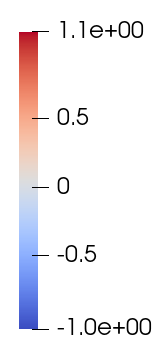}
          &
          \includegraphics[width=0.17\textwidth]{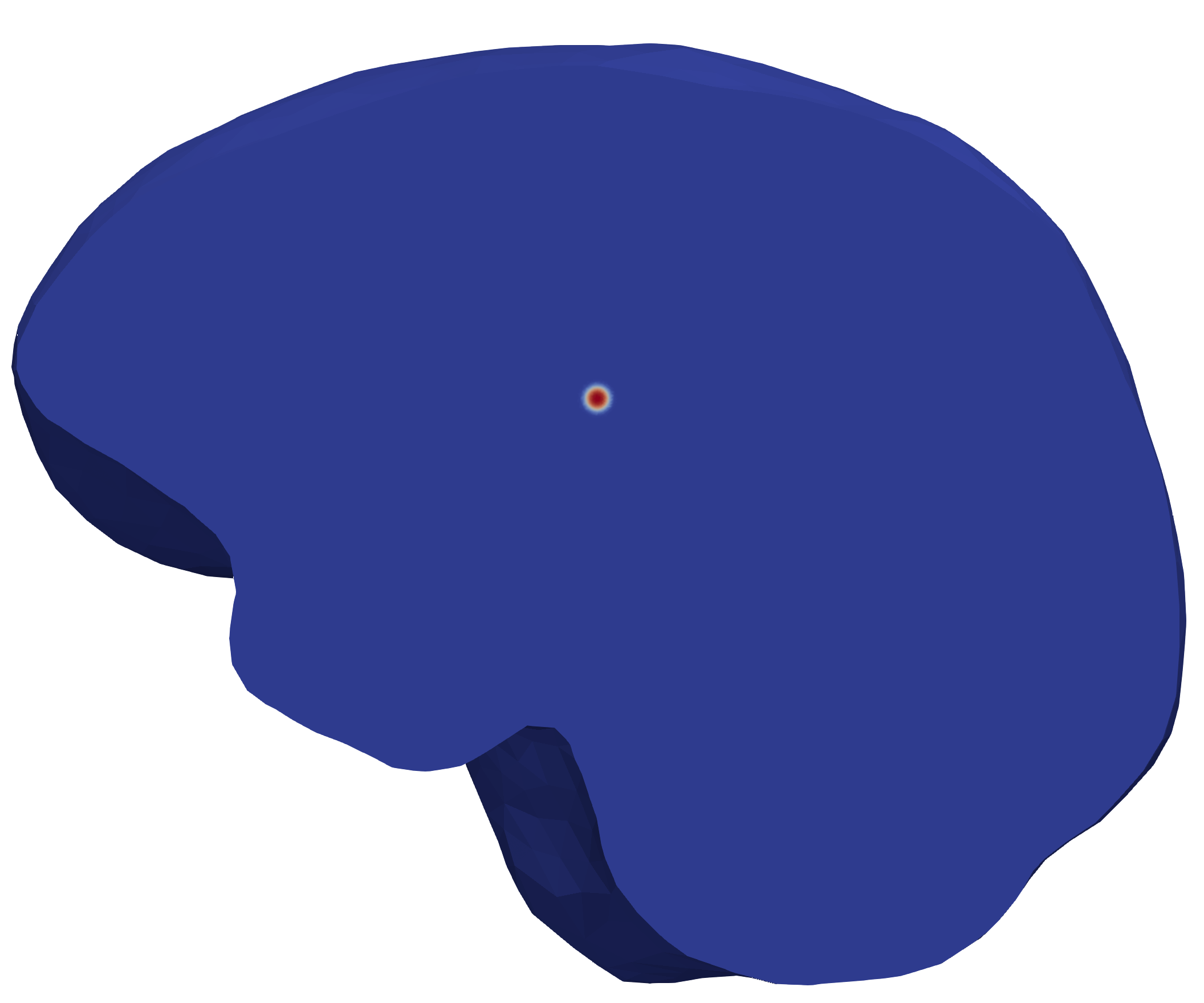}
          & \includegraphics[width=0.17\textwidth]{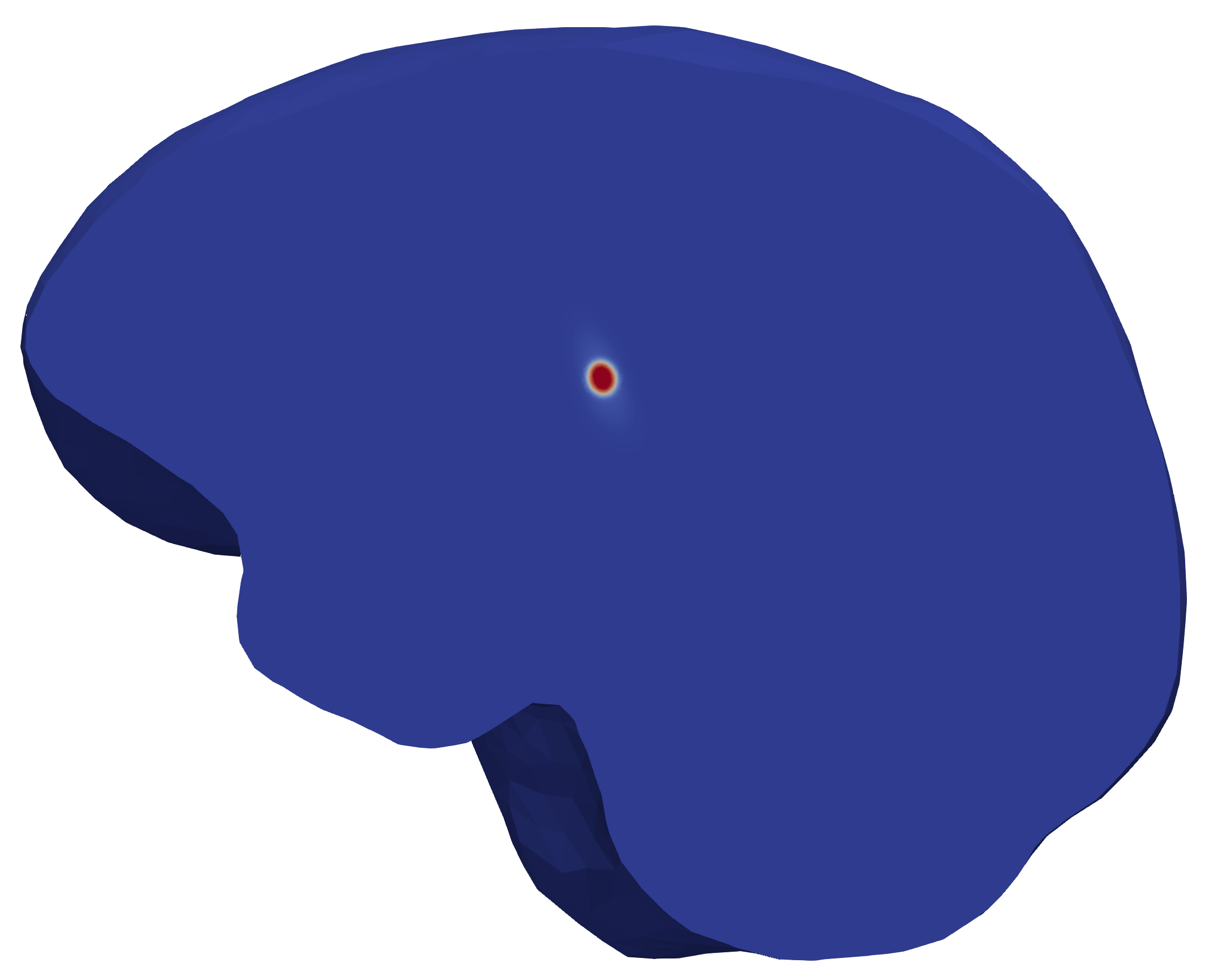}
          &\includegraphics[width=0.17\textwidth]{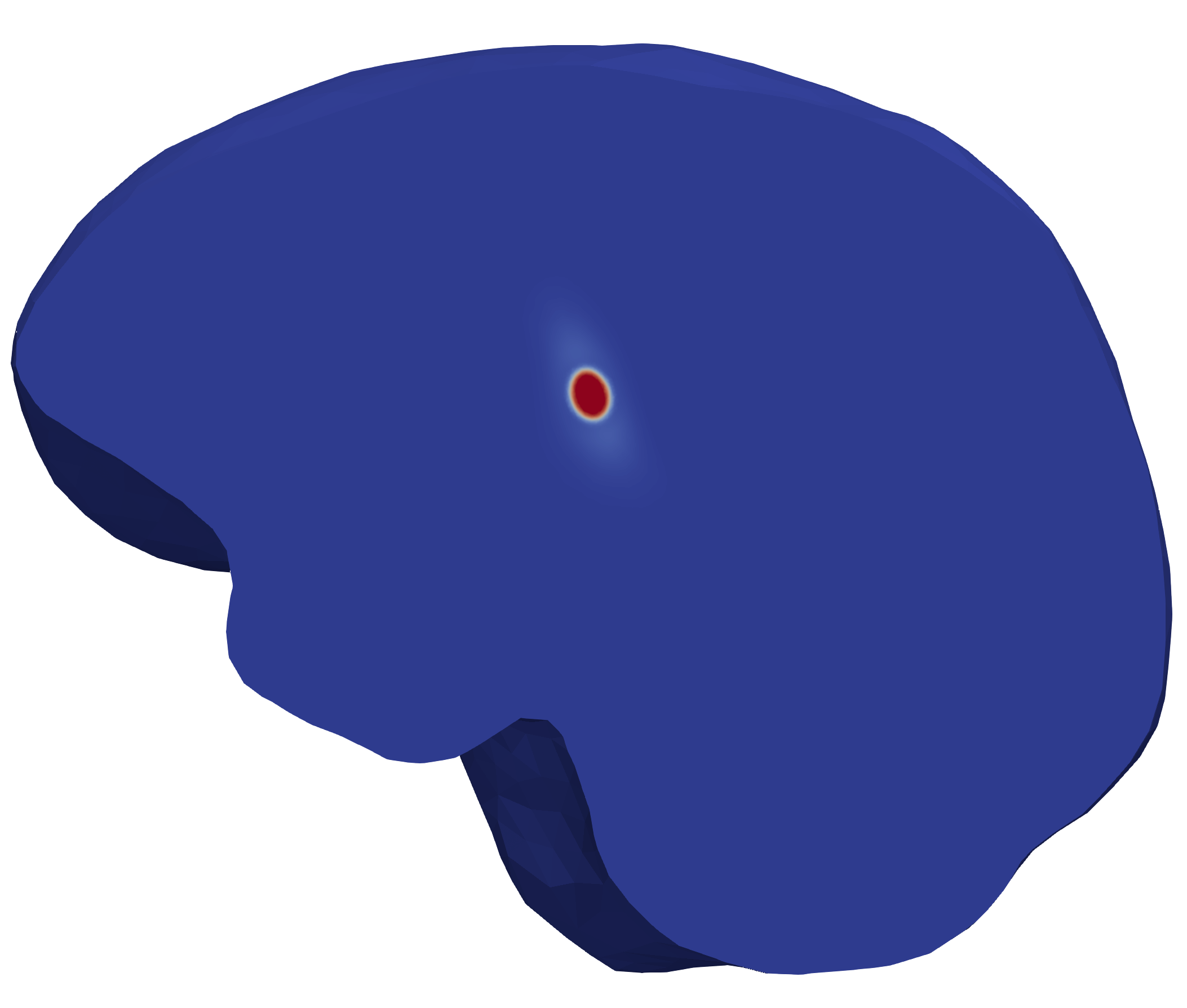}  \\

          \hline
          & & POD-Galerkin & \\
          \includegraphics[width=0.07\textwidth]{legend.png}
          &
          \includegraphics[width=0.17\textwidth]{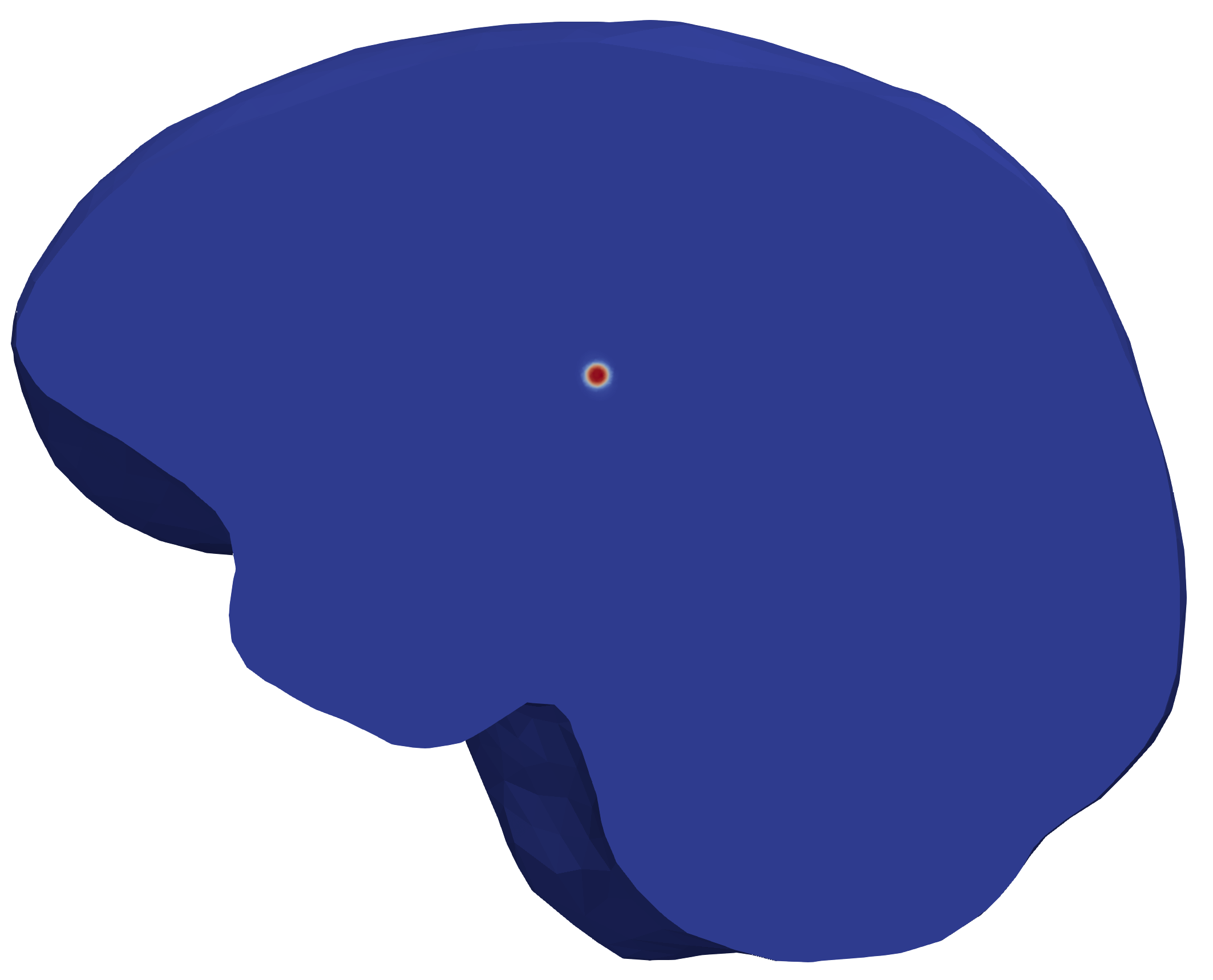}
          & \includegraphics[width=0.17\textwidth]{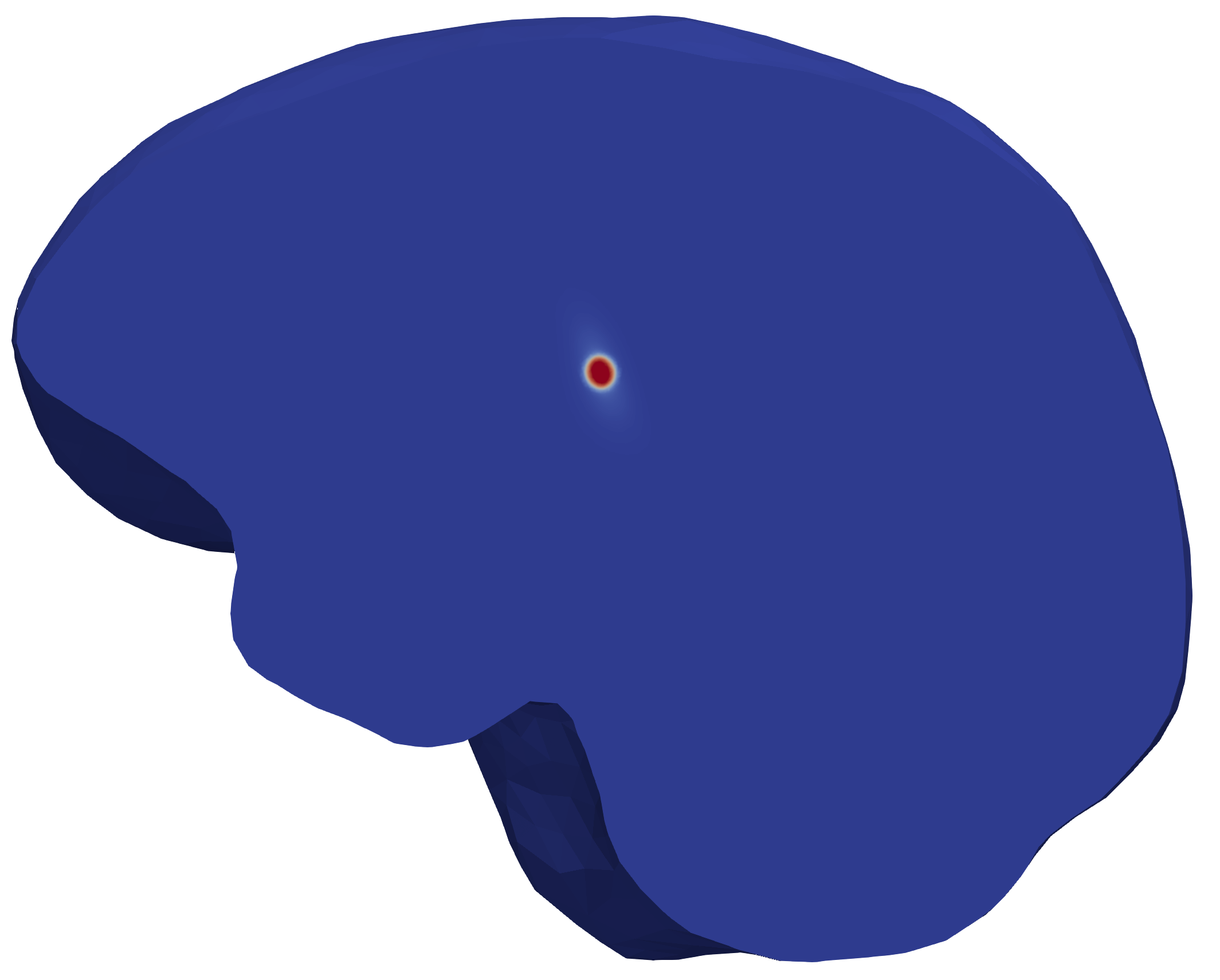}
          &\includegraphics[width=0.17\textwidth]{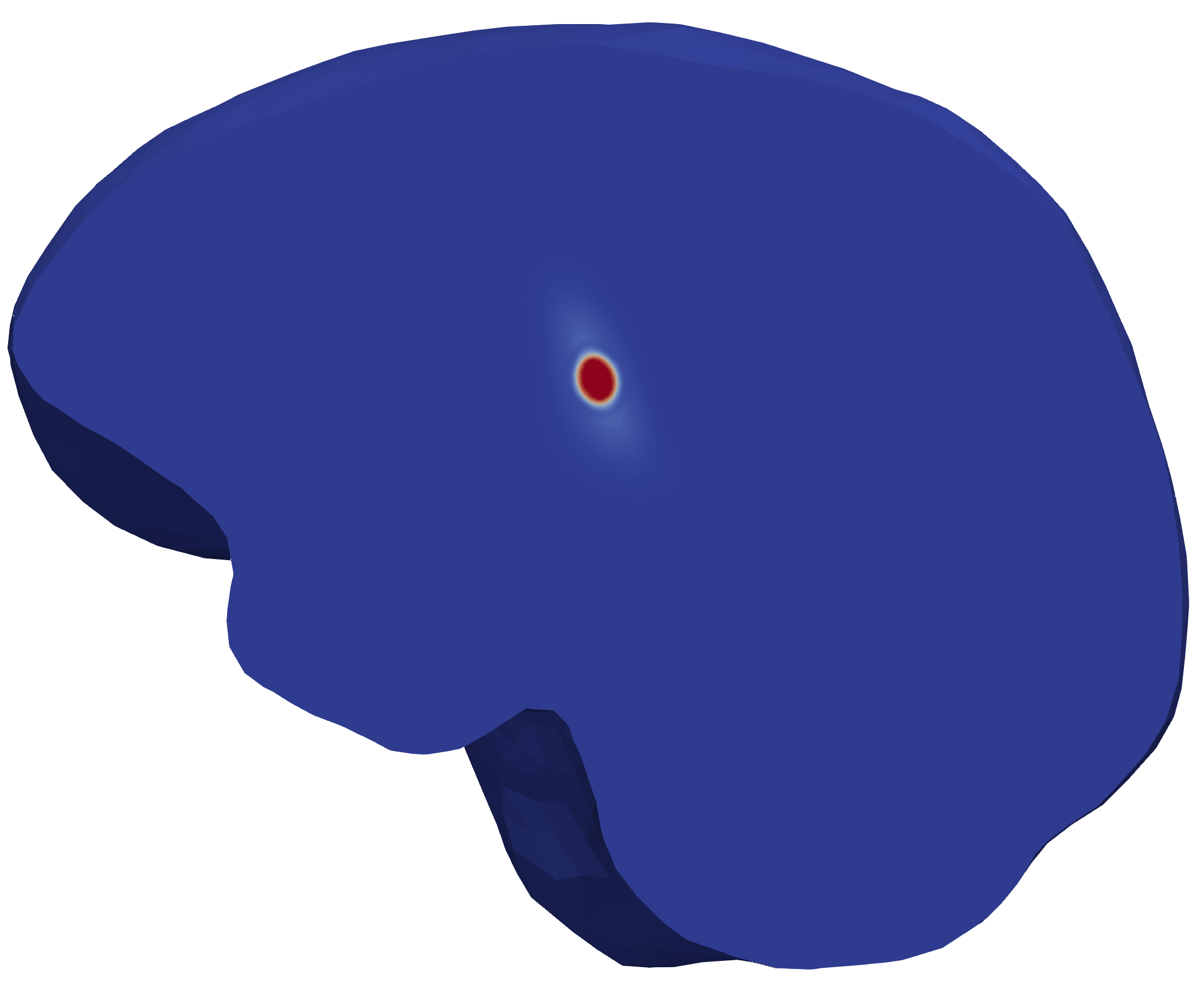}  \\
          \hline
          & & POD-NN& \\
          \includegraphics[width=0.07\textwidth]{legend.png}
          &
          \includegraphics[width=0.17\textwidth]{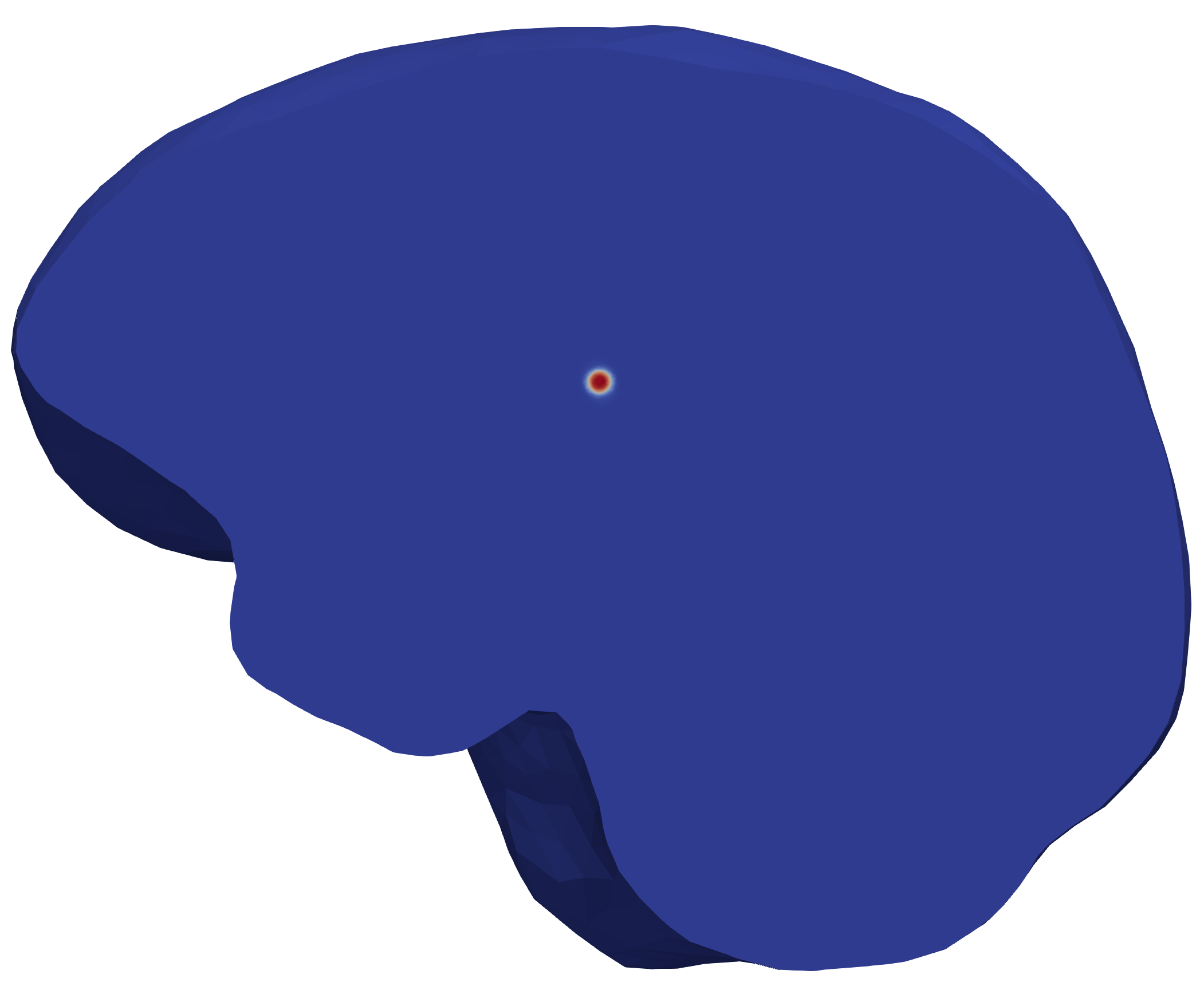}
          & \includegraphics[width=0.17\textwidth]{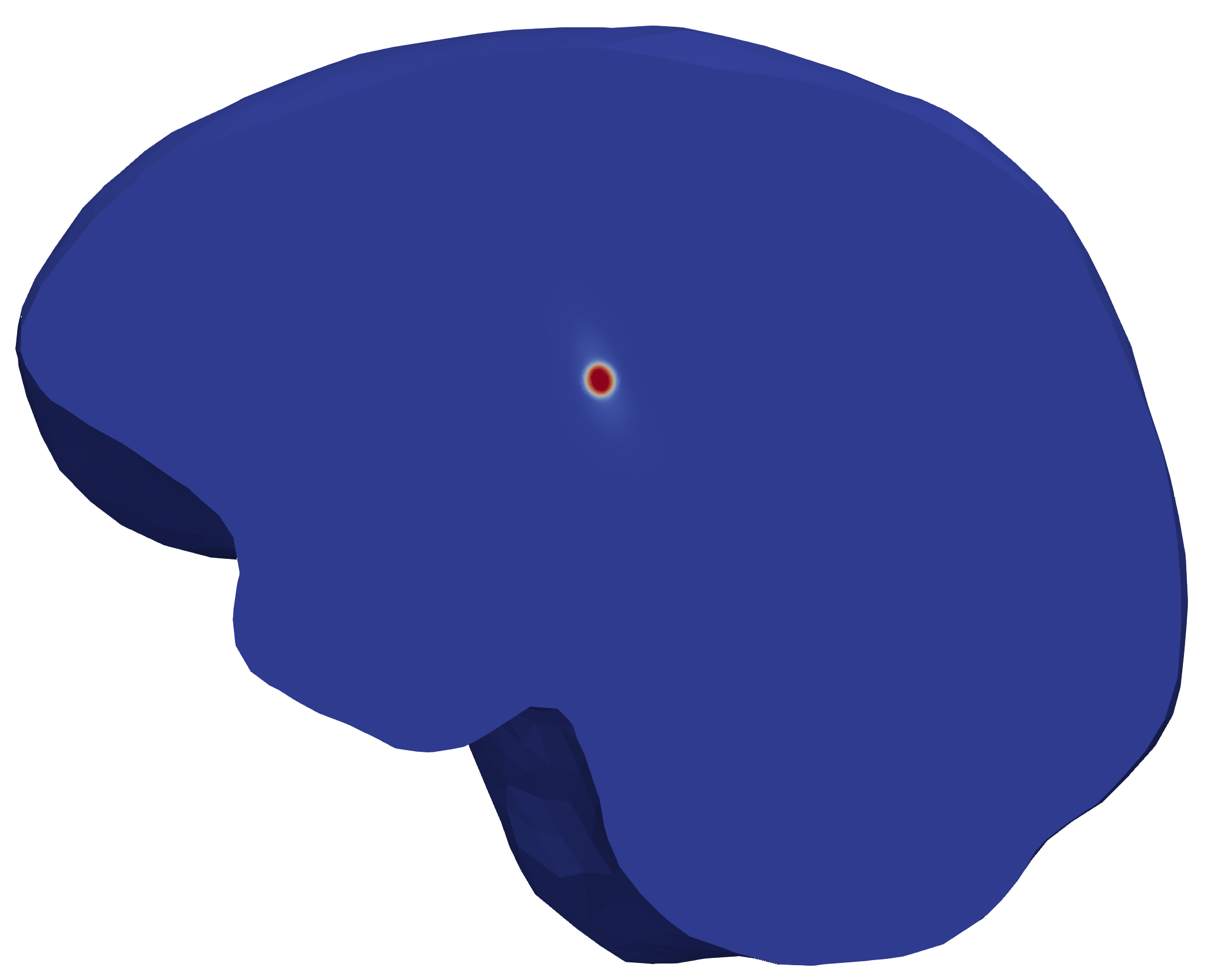}
          &\includegraphics[width=0.17\textwidth]{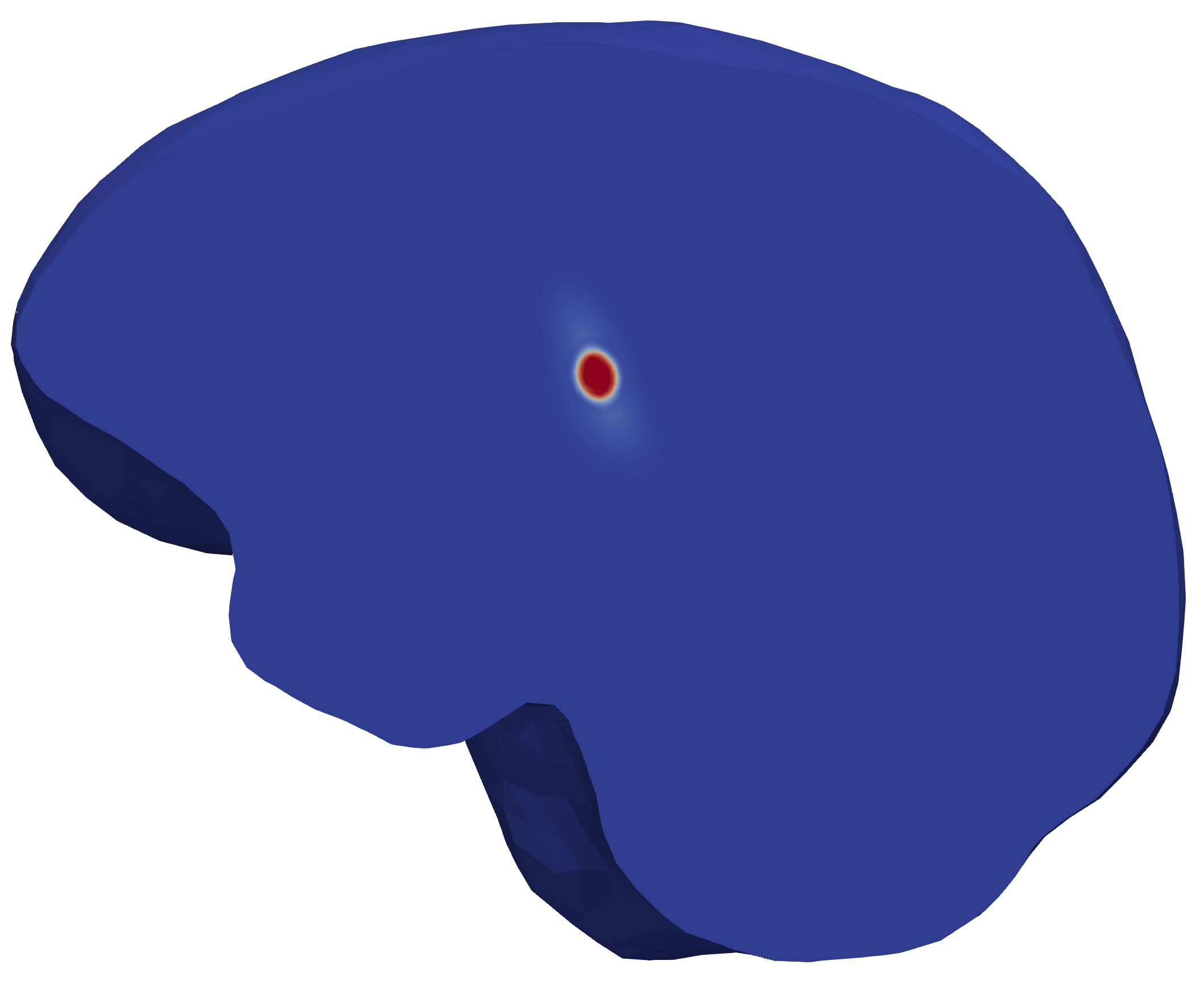}  \\
          \hline
          & & Parameter Estimation & \\
          \includegraphics[width=0.07\textwidth]{legend.png}
          &
          \includegraphics[width=0.17\textwidth]{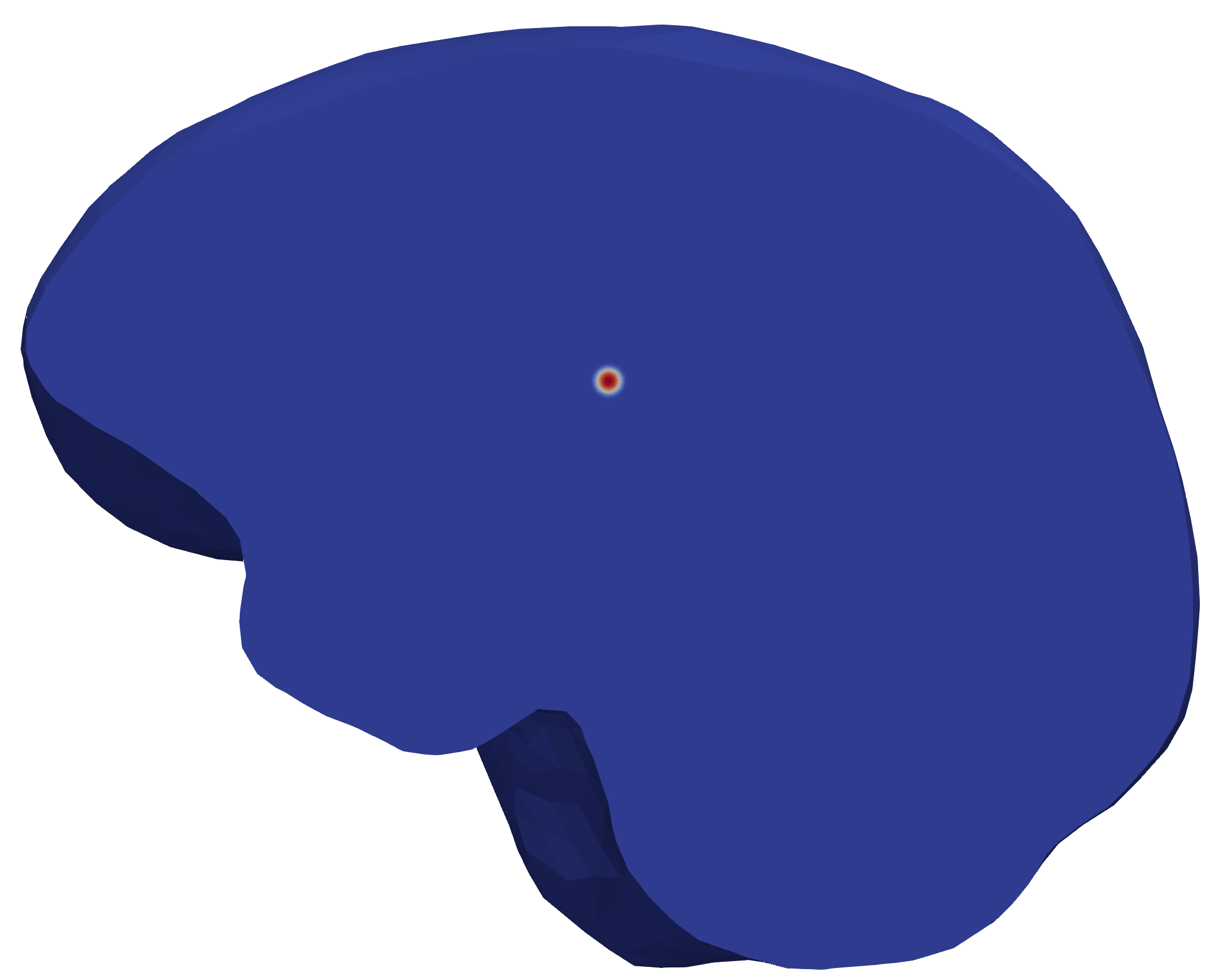}
          & \includegraphics[width=0.17\textwidth]{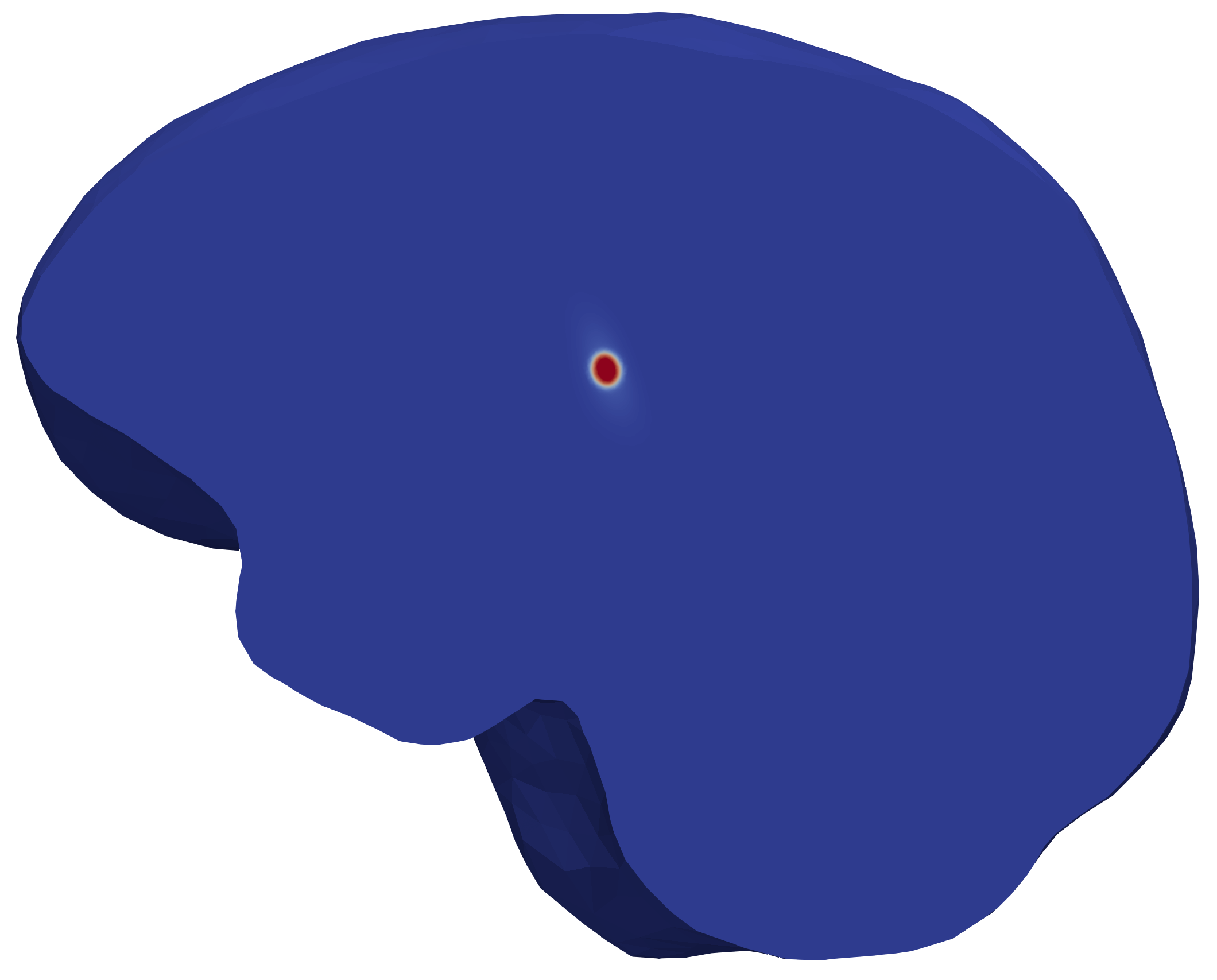}
          &\includegraphics[width=0.17\textwidth]{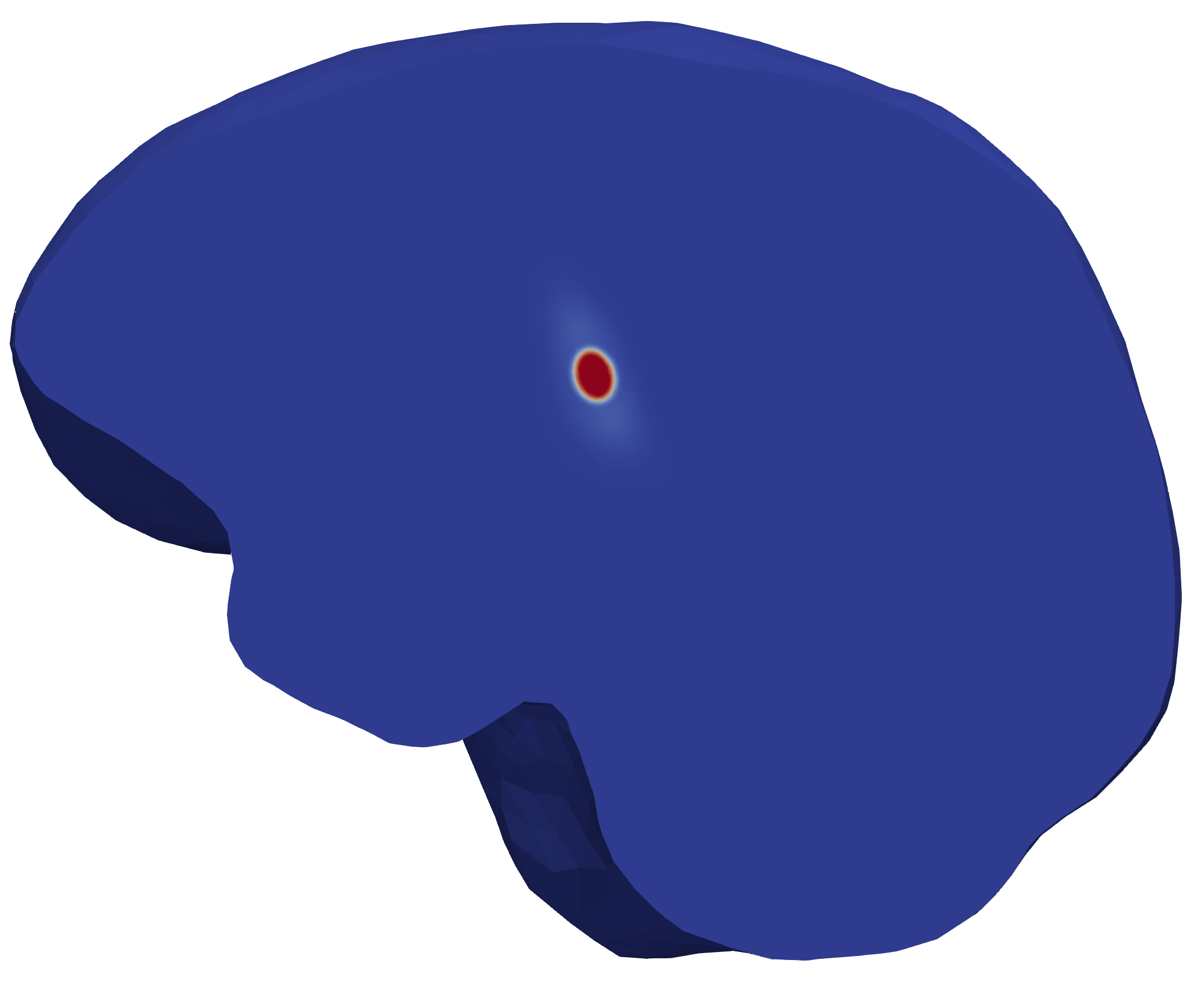} \\
          \hline
        \end{tabular}
      \end{center}

      \begin{minipage}[!t]{0.3\textwidth}
        \begin{center}
          \begin{tabular}{ | c c |}
            \hline
            \textbf{Method} & \textbf{Elapsed Time} \\
            \hline
            FOM & 780 s \\
            \hline
            POD-Galerkin & 475  s \\
            \hline
            POD-NN & 5 s \\
            \hline
          \end{tabular}
        \end{center}
      \end{minipage}
      \qquad
      \begin{minipage}[!t]{0.45\textwidth}
        \begin{center}
          \textbf{Tumor volume}\\
          \includegraphics[width=0.9\textwidth]{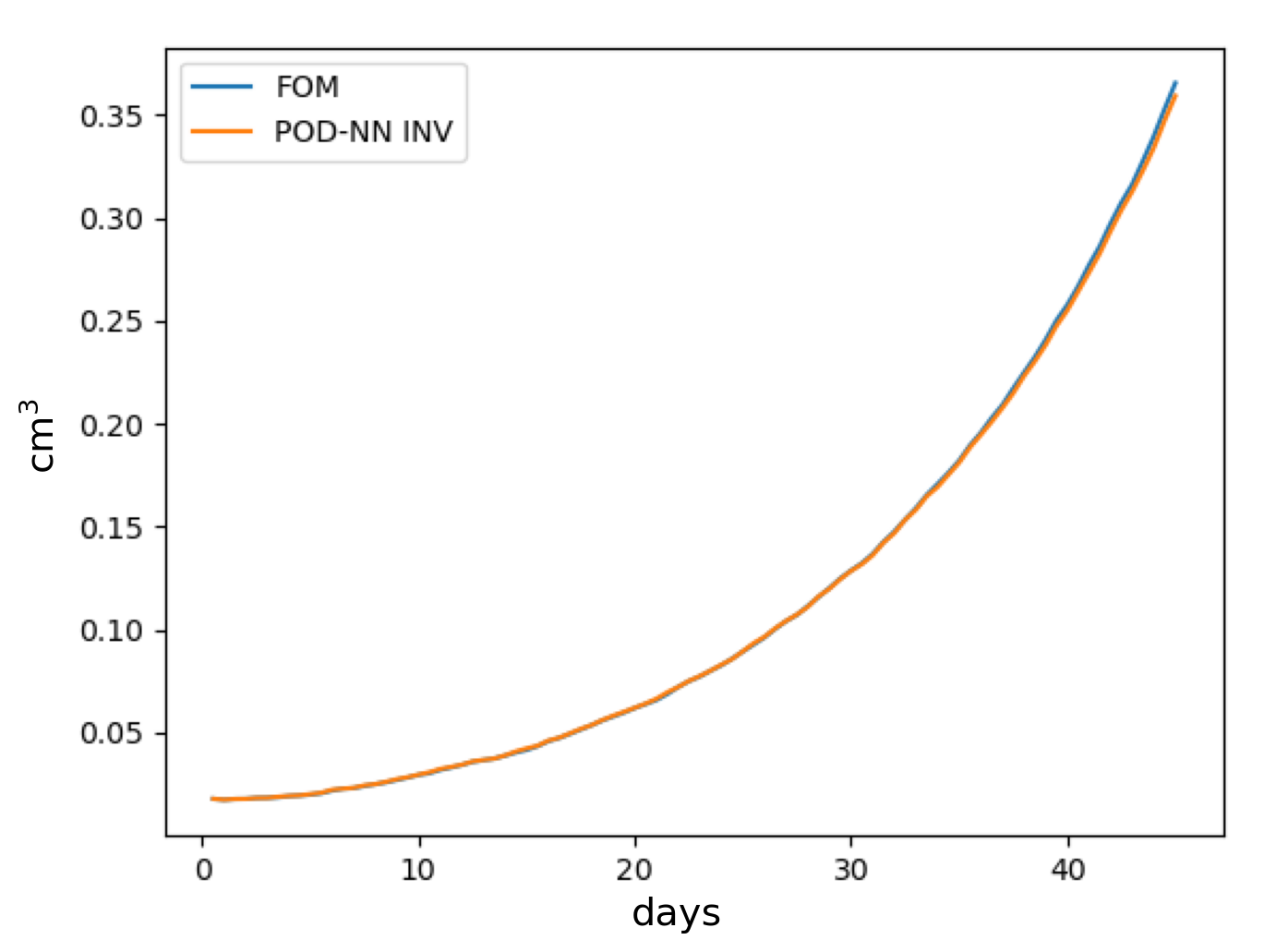}
        \end{center}
      \end{minipage}

      \caption{FOM, POD-Galerkin, POD-NN, and patient-specific FOM solutions  of a GBL concentration  $\phi$ at t=\,0,\,15,\,30 days (top) and corresponding  computational times (bottom,left). Volume fraction of tumor over time (bottom, right). The parameters used in the FOM and POD-Galerkin models are given by Eq.~\eqref{eq:param_spec}, while the results obtained with the parameters estimated by the inverse neural network are given by \eqref{eq:param_inverse}. The FOM evolution is given in blue, and the predicted one in orange.}
      \label{fig:comparison}
    \end{figure*}

    The computational demand of the POD-Galerkin solution is generally high due to the absence of hyper-reduction techniques (see Section \ref{sec:reduced_order_model}), and is particularly sensitive to the number of basis functions selected. When a large number of basis functions is required, it may be more efficient to simulate the FOM first and subsequently project onto the reduced basis. However, in our test case, the number of basis functions needed to accurately reconstruct the original solution is limited, making it feasible to use the POD-Galerkin solutions directly to build the data set, as illustrated by the computational times reported in Fig.~\ref{fig:comparison}.
    Once training is complete, the POD-NN achieves a computational speed-up of approximately 150 times compared to the FOM solver, see Fig.~\ref{fig:comparison}.

    The computational requirement of the POD-Galerkin solution is generally high due to the absence of hyper-reduction techniques (see Section \ref{sec:reduced_order_model}), and is particularly sensitive to the number of basis functions selected. When a large number of basis functions is required, it may be more efficient to simulate the FOM first and subsequently project its solution onto the reduced basis. However, in our test case, the number of basis functions needed to accurately reconstruct the original solution is rather limited, making it feasible to use the POD-Galerkin solutions directly to build the data set, as illustrated by the computational times reported in Fig.~\ref{fig:comparison}.
    Once training is complete, the POD-NN achieves a computational speed-up of approximately 150 times compared to the FOM solver, see Fig.~\ref{fig:comparison}.

    Given the distributions of the tumor at two sufficiently distant instants of time, in order to discriminate between different possible evolutive scenarios with more accuracy, we can produce a data set whose input-output pairs are formed by the vector containing the coefficients of the projections of the tumor distribution over the RB and the patient-specific parameters. To train the inverse neural network, denoted by $\vect{NN}_\text{inv}$, we extract twenty pairs of tumor distributions, each separated by a distance of twenty days, for each of the 750 parameter sets. This results in a total of $N^\text{inv}_\text{data} = 15000$ input-output pairs. These are then split into a training set containing $N^\text{inv}_\text{train} = 11000$ elements and a test set with $N^\text{inv}_\text{test} = 4000$ elements. The mean squared errors over epochs for the neural network $\vect{NN}_\text{inv}$, computed for normalized over the biological range parameters, are shown in Fig.~\ref{fig:err_inv}. Although this result appears to be non-optimal in order to catch the exact parameter of
    a patient (the computed error is about 15\%, see the bottom panel of Fig.~\ref{fig:err_inv}), the simulations performed show that the specific behaviour is actually well captured.

    Giving as input the distribution of the tumor starting from the parameters
    \begin{equation}
      \label{eq:param_spec}
      \begin{aligned}
        & M_0 = \SI{3860.7}{\pascal\day\per\milli\meter\squared}, && \delta_n = \SI{21041}{\per\day},\\
        & \nu = \SI{0.356}{\per\day}, && \kappa = \SI{700.4}{\pascal},\\
        & S_n = \SI{41978}{\per\day}, && \delta = \SI{0.24}{}.
      \end{aligned}
    \end{equation}
    we obtain the following result
    \begin{equation}
      \label{eq:param_inverse}
      \begin{aligned}
        & M_0 = \SI{3924.9}{\pascal\day\per\milli\meter\squared}, && \delta_n = \SI{21922}{\per\day},\\
        & \nu = \SI{0.366}{\per\day}, &&\kappa = \SI{717.5}{\pascal},\\
        & S_n = \SI{43034}{\per\day}, &&\delta = \SI{0.23}{}.\\
      \end{aligned}
    \end{equation}
    In Fig.~\ref{fig:comparison} (top)
    the evolution of the tumor with the actual set of parameters and the evolution with the predicted set is exhibited.
    As we can see in Fig.~\ref{fig:comparison} (right, bottom), the volume fraction is well-tracked over time entailing a good estimation both in terms of tumor morphology.
    The elapsed time for the estimation of the parameters is of the order of seconds (Fig.~\ref{fig:comparison}) since it only requires the evaluation of the trained map at a specific point given by the projected tumor distributions onto the reduced basis.

    \subsection{Global sensitivity analysis}
    To assess the sensitivity of our computational model to variations in input parameters, we performed a global Morris sensitivity analysis on the direct problem \cite{morris1991factorial,saltelli2008global}, using the algorithms recently developed for similar studies \cite{Possenti20201215,vitullo2023sensitivity}. The Morris method is an efficient screening technique that quantifies the influence of each parameter by systematically perturbing them within the input space. This approach allows us to identify the most influential parameters and detect possible interactions or nonlinear dependencies.

    In our study, we generated $r=60$ independent trajectories, each consisting of a unique sequence of input parameter perturbations. By varying one parameter at a time while keeping the others fixed, we computed the elementary effects of each parameter across multiple sampled points in the parameter space. This procedure provides two key sensitivity indices: $\mu^*$, representing the mean absolute elementary effect and quantifying the overall influence of a parameter on the output, and $\sigma$, which captures the variability of elementary effects and serves as an indicator of parameter interactions or nonlinear effects.

\begin{figure}[t!]
      \centering
      {%
      \includegraphics[width=\columnwidth]{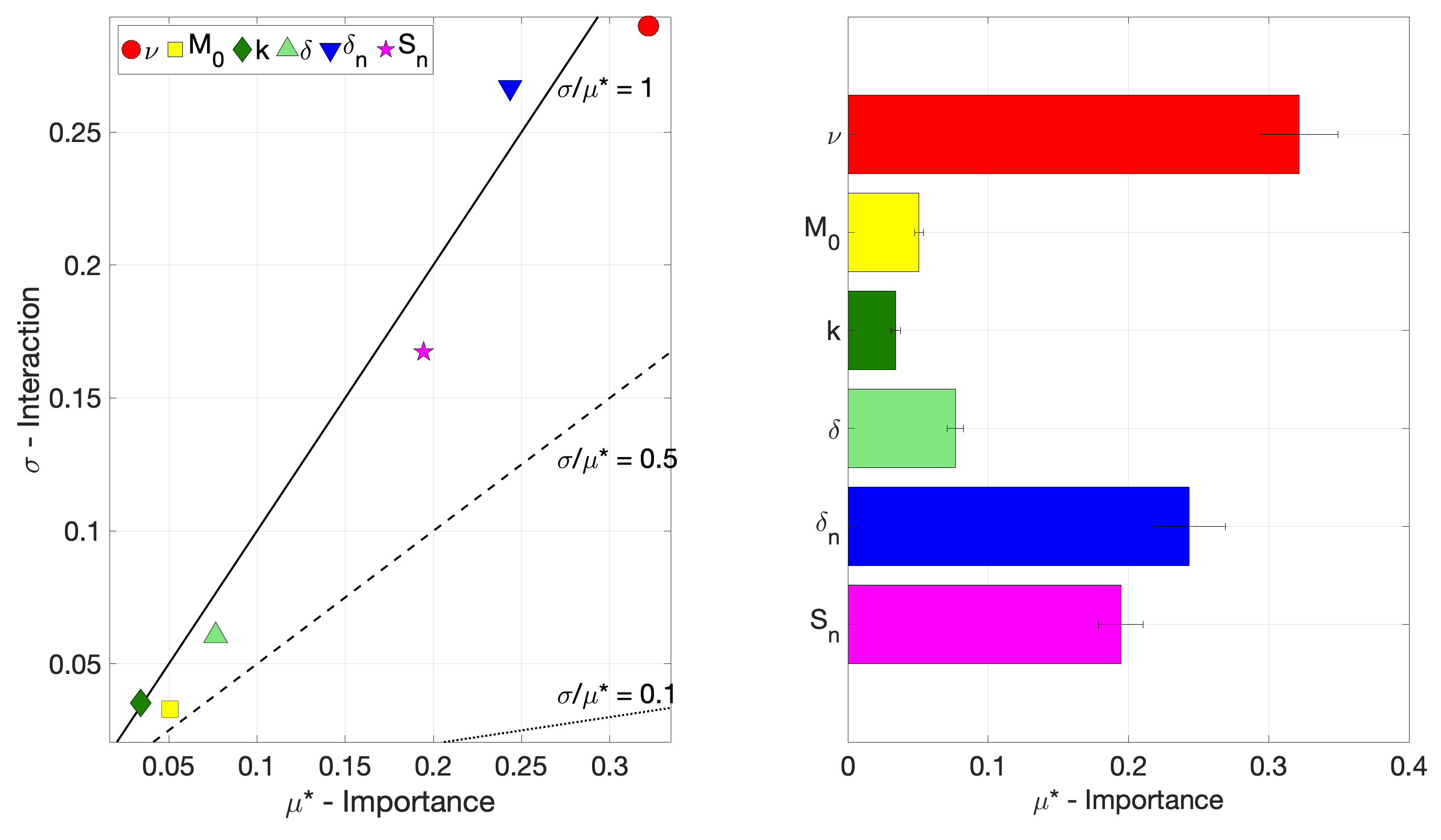}}
      \caption{Global sensitivity analysis results. Distributions of parameters on the importance $\mu^*$ versus interaction $\sigma$ plane (left). The three lines represent different ratios of $\sigma$ and $\mu^*$.
      Histogram of parameter global sensitivity with error bars representing 0.95 confidence interval level of $\mu^*$ (right).}
      \label{fig:MorrisSA}
    \end{figure}

    The output quantity of interest is the total tumor volume at the end of the simulation ($t = 30$ days). The results of the Morris sensitivity analysis are summarized in Figure~\ref{fig:MorrisSA}, where the indices $\mu^*$ and $\sigma$ are reported for each of the six model parameters.

    Our analysis identifies the tumor proliferation rate $\nu$ as the most influential parameter, which aligns with its fundamental role in driving tumor growth dynamics. The oxygen consumption rate $\delta_n$ is also a key determinant, as it directly affects the tumor’s metabolic demand for oxygen. Additionally, the oxygen supply rate $S_n$ exhibits a moderate influence on tumor volume. The remaining three parameters have a negligible direct impact on the output. Notably, both $\delta_n$ and $S_n$ exhibit high $\sigma$ values, suggesting a strong interaction between them, as they jointly regulate the tumor’s nutrient availability and metabolic balance.

    These findings highlight the dominant role of tumor proliferation and oxygen dynamics in glioblastoma growth, reinforcing the importance of accurately estimating these parameters in patient-specific predictive modeling.

    \subsection{Local sensitivity analysis}

    To evaluate the robustness of the inverse problem solution with respect to small variations in input data, we performed a local sensitivity analysis using a Monte Carlo approach. Specifically, we assessed how minor perturbations in the input coefficients of $\vect{NN}_\text{inv}$ affect the estimated model parameters.

    We considered the reference parameter set defined in \eqref{eq:param_spec}, along with the corresponding tumor distributions projected onto the reduced-order basis at two distinct time points ($t=9.5$ days and $t=29.5$ days). The projection coefficients of these tumor distributions serve as inputs to $\vect{NN}_\text{inv}$. To simulate measurement uncertainty, we introduced a $1\%$ Gaussian random perturbation to these coefficients.

\begin{figure}[t!]
      \centering
      {%
      \includegraphics[width=\columnwidth]{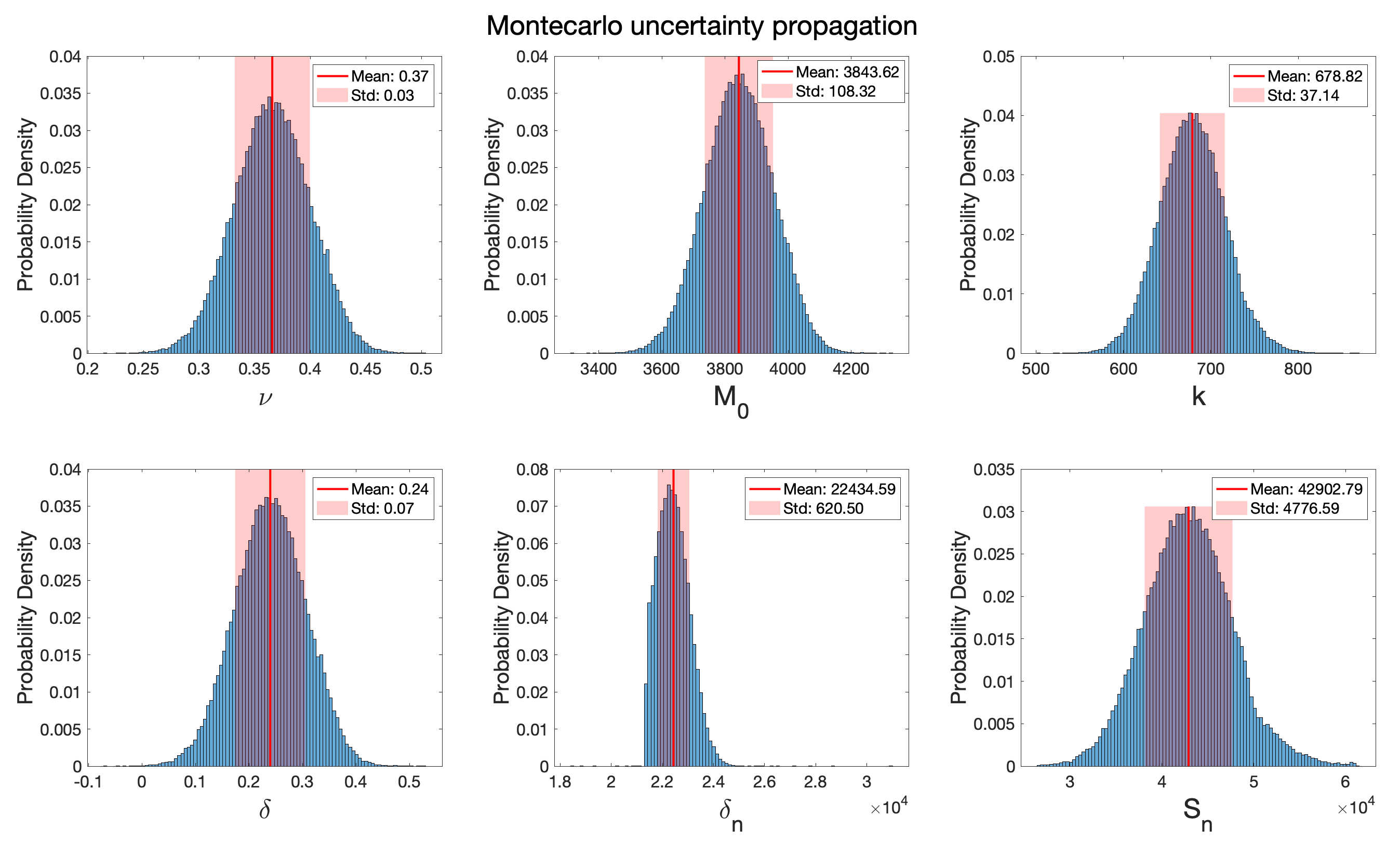}}
      \caption{Local sensitivity analysis results: probability distributions for each of the parameter estimated as an output of the inverse problem. For each distribution, the corresponding mean and standard deviation values are also displayed.}
      \label{fig:montecarloSA}
    \end{figure}

    A total of $N=\SI{100000}{}$ Monte Carlo samples were generated, and the resulting parameter estimates were analyzed by examining their probability distributions, as shown in Figure~\ref{fig:montecarloSA}.

    The estimated parameters exhibit approximately normal distributions, with mean values closely matching those predicted by $\vect{NN}_\text{inv}$ in the absence of input perturbations. Among all parameters, the hypoxia threshold $\delta$ demonstrates the highest standard deviation relative to its biological range, followed by the Young’s modulus of the brain $k$. Interestingly, the parameters $\delta$ and $\kappa$, while having low influence on the model outputs as shown by global sensitivity analysis, exhibit the largest uncertainty in the inverse problem. This is consistent with the fact that less influential parameters are weakly constrained by the data, leading to higher variability in their inferred values. This observation suggests that model calibration efforts should focus on the most influential parameters (e.g., $\nu$, $\delta_n$) to ensure robustness.

    These findings underscore the robustness of the inverse problem formulation and the effectiveness of the reduced-order model in ensuring stable parameter estimates, even in the presence of small input uncertainties.

    \section{Conclusions}
    In this work, we presented a mechanistic learning framework for patient-specific prediction of GBL growth, combining a physics-based diffuse-interface model with a reduced-order representation via POD and neural networks. This hybrid approach enables the fast and accurate estimation of key biological parameters from longitudinal imaging data, with a computational speed-up of approximately 99\% and an accuracy of 96\% in forecasting tumor volume (see Fig.~\ref{fig:comparison}). Once trained, the surrogate model allows real-time simulation of tumor evolution, offering a practical solution for time-sensitive clinical scenarios.

    The framework effectively addresses both the direct and inverse problems of GBL modeling. Through a global Morris sensitivity analysis, we identified the key parameters driving tumor growth dynamics and their interactions, ensuring interpretability and model parsimony. Complementarily, the local Monte Carlo sensitivity analysis confirmed the robustness of the inverse mapping under small perturbations in the input data, highlighting the reliability of parameter estimates for clinically relevant quantities. Notably, discrepancies in less influential parameters (e.g., hypoxia threshold, tissue stiffness) showed limited impact on the predictive accuracy, reinforcing the consistency of the reduced-order inversion scheme.

    While the current study is based on synthetic data, its structure lays the mathematical foundation for clinical translation. The model formulation, numerical reduction, and learning architecture were designed to be modular and scalable. Future developments will focus on extending the model to include treatment effects and phenotype heterogeneity \cite{celora2021phenotypic}, as well as on coupling tumor mechanics and growth \cite{ambrosi2002mechanics,jain2014role,riccobelli2024elastocapillarity}. Efforts will also be devoted to improving generalization across varying anatomical geometries and initial conditions without the need for retraining \cite{lassila2010parametric,manzoni2012model,zhang2024shape}.

    Hence, the current contribution is intentionally focused on the methodological innovation, providing a rigorous computational foundation. The application to real patient data is the object of an ongoing clinical study, which will be addressed in a follow-up publication aimed at demonstrating translational relevance in neuro-oncology.

    In conclusion, this study represents a significant step towards the development of interpretable, scalable, and efficient computational tools for precision medicine in the management of GBL.

    \appendix
    \section{Appendix: details of the medical protocol and neuroimaging techniques}
    \subsection{Clinical protocol}
    \label{appendix_protocol}
    This study is part of a collaboration program between Foundation IRCCS Neurological Institute Carlo Besta, Department of Neurosurgery and Neuroradiology, and Politecnico di Milano, MOX – Modeling and Scientific Computing, Department of Mathematics.
    At a hospital stage, we started a prospective observational trial, named GLIOMATH (GLIO\-blastoma MATHematics), enrolling patient with GBL submitted to surgical removal or biopsy, adjuvant therapy and follow-up based on normal clinical practice, in which specific MRI data for each patient were used as input data building a personalized virtual environment. The study was conducted according to the guidelines of the Declaration of Helsinki, and approved by the Ethics Committee of Fondazione Istituto Neurologico Carlo Besta (protocol code GlioMath, nr. 49/2016; date of approval: 13 July 2016).
    Patients older than 18 years old with suspected, newly diagnosed, untreated GBL and eligible for surgical removal or biopsy of their lesion were considered for participating in our trial; exclusion criteria were inability to give consent due to cognitive deficits or language disorders, or, for women, pregnancy or lactation. The patients were enrolled in a prospective observational study; the evaluation was based on the normal clinical practice \cite{Acerbi_2020}. All patients underwent neurological examination, preoperative volumetric MRI including DTI (3 Tesla MRI scan – Philips), and recording of concomitant medications. Patients were scheduled for surgical removal or biopsy as judged by the surgeon; in both cases, the procedures were performed in a standard manner, with any surgical tools as preferred by the operating surgeon, and neurophysiological monitoring when necessary. The histopathological and molecular analysis of the tumor samples were performed according to the 2016 or 2021 WHO classification of CNS tumors \cite{Sanai_2009}.
    Clinical and radiological post-operative examination were carried out the usual institutional practice. The early clinical evaluation included neurological examination and volumetric contrast-enhanced MRI for estimation of extent of resection, within 72 hours from the surgical intervention. The protocol for early postoperative MRI was the same as performed in preoperative setting without the DTI, that was excluded due to the possibility of artifacts caused by the presence of air in the surgical cavity; the following radiological exams, performed every two months, were performed according to the same protocol of preoperative MRI with DTI.
    All patients, upon confirmation of histologic diagnosis of GBL, were offered adjuvant radio- and chemotherapy, according to the Stupp protocol and tailored on the basis of patient age, performance status and methylation status of MGMT gene promoter, according to the EANO guideline \cite{Weller_2017}.
    The surgical and trial databases of the above mentioned study have been collected anonymously for the scientific purposes; written informed consent was obtained for each case. Exclusively anonymized neuroradiological data were employed for the secondary phase of the study consisting in developing of a multi-scale mathematical model and simulating GBL invasion from the patient-specific data collected from MRI studies.

    \subsection{Neuroimaging acquisition and segmentation}
    \label{appendix_DT}
    The radiological protocol included volumetric axial whole brain T1-weighted MRI at $1\,mm \times 1\,mm \times 1\,mm$ spatial resolution and volumetric axial whole brain T1-weighted MRI at same spatial resolution after paramagnetic contrast administration, useful for illustrating the structural anatomy of the patient’s brain and to calculate the total volume of tumor extension after segmentation procedure; axial whole brain 3D-FLAIR image at 1 mm × 1 mm × 1 mm spatial resolution, useful to delineate the outline of the tumor and peri-tumor rim by suppressing signal from cerebrospinal fluid. A set of 147 diffusion-weighted images DTI at 2 mm × 2 mm × 2 mm spatial resolution with anterior–posterior phase encoding direction with different b-value was finally acquired; all diffusion-sensitising directions were sampled uniformly on the hemisphere and an additional B0 image was acquired with reversed phase encoding direction, as posterior-anterior encoding, for helping in geometric distortion correction. The images obtained through MRI are segmented using the software 3D Slicer and the mesh is generated using the VMTK library. Finally, the DTI data are then analyzed using the library ANIMA\footnote{\url{https://github.com/Inria-Empenn/Anima-Public}} to reconstruct the tensors $\mathsf{D}$ and $\mathsf{T}$.
    Specifically, the six independent components of the diffusion tensor $\mathsf{D}$ can be directly derived from DTI images as explained above, while the anisotropy tensor $\mathsf{T}$ is created from the components of $\mathsf{D}$. Indeed, the tensor $\mathsf{T}$ can be parametrized by a tuning parameters to modulate the anisotropy of DTI, as done in \cite{ZAMM18}. The tensor $\mathsf{D}$ can be written as:
    \begin{equation*}
      \mathsf{D} = \lambda_1 \vect{e}_1 \otimes \vect{e}_2 + \lambda_2 \vect{e}_2 \otimes \vect{e}_2 + \lambda_3 \vect{e}_3 \otimes \vect{e}_3,
    \end{equation*}
    where $\lambda_i$ and $\vect{e}_i$ for $i=1,2,3$ are the descending order eigenvalues and the corresponding eigenvectors of $\mathsf{D}$. Since $\mathsf{T}$ has the same eigenvalues of $\mathsf{D}$, we set:
    \begin{align*}
      & \hat{\mathsf{T}} =
      \begin{aligned}[t]
        & a_1(r) \lambda_1 \vect{e}_1 \otimes \vect{e}_2
        + a_2(r) \lambda_2 \vect{e}_2 \otimes \vect{e}_2 + \\
        & + a_3(r) \lambda_3 \vect{e}_3 \otimes \vect{e}_3,
      \end{aligned} \\
      & \mathsf{T} = \frac{3}{\mathrm{tr}(\hat{\mathsf{T}})}\hat{\mathsf{T}}
      = \frac{3}{a_1(r) \lambda_1 + a_2(r) \lambda_2 + a_3(r) \lambda_3} \hat{\mathsf{T}}.
    \end{align*}
    where $a_i(r)$ for $i=1,2,3$ are functions of the anisotropy controlling factor $r$ of the form:
    \begin{equation}
      \begin{bmatrix}
        a_1(r) \\
        a_2(r) \\
        a_3(r)
      \end{bmatrix}
      =
      \begin{bmatrix}
        r & r & 1 \\
        1 & r & 1 \\
        1 & 1 & 1 \\
      \end{bmatrix}
      \begin{bmatrix}
        c_l \\ c_p \\ c_s
      \end{bmatrix}
      \label{eq:matrix}
    \end{equation}
    In \eqref{eq:matrix}, $c_l$, $c_p$ and $c_s$ are defined as:
    \begin{align*}
      & c_l = \frac{\lambda_1 - \lambda_2}{\lambda_1 + \lambda_2 + \lambda_3}, \\
      & c_p = \frac{2(\lambda_2 - \lambda_3)}{\lambda_1 + \lambda_2 + \lambda_3}, \\
      & c_s = \frac{3\lambda_3}{\lambda_1 + \lambda_2 + \lambda_3},
    \end{align*}
    and they are the linear, planar and spherical anisotropy coefficients, respectively. Notice that when $r=1$ the anisotropy is not emphasized, and the tensor $\mathsf{T}$ is given by a simple re-scaling of $\mathsf{D}$.
    In this work, we set $r=3$.

    \section*{Acknowledgements}
    This work was partly supported by MUR  through the grants  PRIN 2020 Research Project MATH4I4 and Dipartimento di Eccellenza 2023-2027.
    DR and FB have been partially supported by the PRIN 2022 project \emph{Mathematical models for viscoelastic biological matter}, Prot. 202249PF73 -- Funded by European Union - Next Generation EU -- Italian Recovery and Resilience Plan (PNRR) - M4C1, CUP D53D23005610001 for DR, CUP J53D23003590008 for FB. PV and PZ acknowledge the support of the grant MUR PRIN 2022 No. 2022WKWZA8 Immersed methods for multiscale and multiphysics problems (IMMEDIATE) funded by the Next Generation EU Program, Mission 4, Comp. 2, CUP D53D23006010006.
    DR, SG, and PC are members of \emph{Gruppo Nazionale di Fisica Matematica} (GNFM), while PV, FB, AM, and PZ are members of \emph{Gruppo Nazionale per il Calcolo Scientifico} (GNCS) of Istituto Nazionale di Alta Matematica (INdAM).
    \printbibliography

    \end{document}